\newif\ifsingle
\singletrue 

\newif\ifFullVersion
\FullVersiontrue 

\ifsingle
\pdfoutput=1
\documentclass[12pt,draftclsnofoot, onecolumn]{IEEEtran}		
\else		
\pdfoutput=1
\documentclass[10pt,final, twocolumn]{IEEEtran}
\fi

\usepackage{times}
\usepackage{amsmath,dsfont}
\usepackage{amssymb,amsthm}
\usepackage{epsfig,verbatim}
\usepackage{subfigure}
\usepackage{setspace}
\usepackage{color}
\usepackage{cite}
\usepackage{epstopdf}
\usepackage{graphics}
\usepackage{accents}
\usepackage{acronym}
\usepackage[bookmarks,colorlinks]{hyperref}
\usepackage{booktabs}
\usepackage{mathtools}
\usepackage{algorithm}
\usepackage{algorithmic}
\usepackage{enumitem}

\newcommand{\myVec}[1]{{\bf {#1}}}
\newcommand{\myMat}[1]{{\bf{#1}}}
\newcommand{\mySet}[1]{\mathcal{#1}}
\newcommand{\E}{\mathds{E}}		 			

\newcommand{\dF}{d_\mathrm{F}}

\newtheorem{theorem}{Theorem}

\newtheorem{lemma}{Lemma}

\ifsingle

\else

\fi 

\acrodef{adc}[ADC]{analog-to-digital convertor}
\acrodef{cs}[CS]{compressed sensing}
\acrodef{csi}[CSI]{channel state information}
\acrodef{dma}[DMA]{dynamic metasurface antenna}
\acrodef{dtft}[DTFT]{discrete-time Fourier transform}
\acrodef{dnn}[DNN]{deep neural network} 
\acrodef{map}[MAP]{maximum a-posteriori probability}
\acrodef{snr}[SNR]{signal-to-noise ratio}
\acrodef{sinr}[SINR]{signal-to-interference-and-noise ratio}
\acrodef{bs}[BS]{base station} 
\acrodef{em}[EM]{electromagnetic} 
\acrodef{iot}[IOT]{Interent of Things}
\acrodef{mimo}[MIMO]{multiple-input multiple-output}
\acrodef{mse}[MSE]{mean-squared error}
\acrodef{pdf}[PDF]{probability density function}
\acrodef{rv}[RV]{random variable}
\acrodef{fec}[FEC]{forward error correction}
\acrodef{lti}[LTI]{linear time-invariant}
\acrodef{wss}[WSS]{wide-sense stationary}
\acrodef{psd}[PSD]{power spectral density}
\acrodef{ris}[RIS]{reconfigurable intelligent surface}
\acrodef{lis}[LIS]{large intelligent surface}
\acrodef{ser}[SER]{symbol error rate} 
\acrodef{ber}[BER]{bit error rate} 
\acrodef{sgd}[SGD]{stochastic gradient descent} 
\acrodef{isi}[ISI]{intersymbol interference}  
\acrodef{awgn}[AWGN]{additive white Gaussian noise} 
\acrodef{ut}[UT]{user terminal} 
\acrodef{mmw}[mmWave]{millimeter wave}
\acrodef{upa}[UPA]{uniform planar array}

\IEEEoverridecommandlockouts

\newcommand{\p}{\mathbf{p}}

\newcommand{\pil}{\mathbf{p}_{i,l}}

\newcommand{\Ail}{{A}_{i,l}}



\title{Beam Focusing for Near-Field Multi-User MIMO Communications 
}
\author{  
	\IEEEauthorblockN{Haiyang Zhang, Nir Shlezinger, Francesco Guidi, Davide Dardari, Mohammadreza F. Imani, and Yonina C. Eldar\\
	} 
	\thanks{
	Parts of this work were accepted for presentation at the 2021 IEEE International Conference on Acoustics, Speech, and Signal Processing (ICASSP) as the paper \cite{zhang2021ICASSP}.
		This project was sponsored in part by  
		the European Union’s H2020 research and innovation program under grant No. 646804-ERC-COG-BNYQ, in part by the Israel Science Foundation under grant No. 0100101, and in part by the Theory Lab, Central Research Institute, 2012 Labs, Huawei Technologies Co.,Ltd.
		H. Zhang and Y. C. Eldar are with the Faculty of Math and CS, Weizmann Institute of Science, Rehovot, Israel (e-mail: \{haiyang.zhang; yonina.eldar\}@weizmann.ac.il). 
		N. Shlezinger is with the School of ECE, Ben-Gurion University of the Negev, Beer-Sheva, Israel (e-mail: nirshl@bgu.ac.il). 
		F. Guidi is with the National Research Council of Italy, Institute of Electronics, Computer and Telecommunication
Engineering, Bologna, Italy (e-mail: francesco.guidi@ieiit.cnr.it). D. Dardari is with the Department of Electrical, Electronic, and Information Engineering “Guglielmo Marconi” - DEI-CNIT,
University of Bologna, 
Cesena, Italy (e-mail:davide.dardari@unibo.it). 
M. F. Imani is with the School of ECEE, Arizona State University, Tempe, AZ (email: mohammadreza.imani@asu.edu).}

	\vspace{-1.0cm}
	
}
\vspace{-0.75cm}

\begin{document}
	
	\maketitle
	\pagestyle{plain}
	\thispagestyle{plain}
\begin{abstract}
Large antenna arrays and high-frequency bands are two key features of future wireless communication systems. The combination of large-scale antennas with high transmission frequencies often results in the communicating devices operating in the near-field (Fresnel) region. In this paper, we study the potential of beam focusing, feasible in near-field operation, in facilitating high-rate multi-user downlink \ac{mimo} systems. As the ability to achieve beam focusing is dictated by the transmit antenna, we study near-field signalling considering different antenna structures, including fully-digital architectures, hybrid phase shifter-based precoders, and the emerging \ac{dma} architecture for massive \ac{mimo} arrays.
We first provide a mathematical model to characterize  near-field wireless channels as well as the transmission pattern for the considered antenna architectures. Then,  we formulate the beam focusing problem for the goal of maximizing the achievable sum-rate in multi-user networks. We propose efficient solutions based on the sum-rate maximization task for fully-digital, (phase shifters based-) hybrid and \ac{dma} architectures.  
 Simulation results show the feasibility of the proposed beam focusing scheme for both single- and multi-user scenarios. In particular, the designed focused beams are such that users residing at the same angular direction can communicate reliably without interfering with each other, which is not achievable using conventional far-field beam steering.

\end{abstract}

\acresetall

\vspace{-0.4cm}
\section{Introduction}
\vspace{-0.1cm}

Wireless communication over high-frequency millimeter wave (mmWave) and terahertz (THz) spectrum is regarded as a key technology for beyond 5G communications, due to its capability of enhancing data-rates thanks to the large available bandwidth.
In order to compensate for the dominant path loss characterizing transmissions in high frequencies, wireless \acp{bs} operating in these bands will be equipped with large antenna arrays \cite{zhang2020prospective}.  
A byproduct of utilizing large-scale antennas is that  high-frequency communication may take place in the near-field (Fresnel) region, 
as opposed to conventional wireless systems, typically operating in the far-field regime. 
More specifically, the near-field distance can be several dozens of meters for relatively small antennas/surfaces at mmWave and THz \cite{guidi2019radio,DarDec:20}. This implies that the far-field model, assuming plane wavefronts of the \ac{em} field rather than spherical ones, no longer holds  at practical distances.
Managing the spherical wavefront of the  signals brings forth the possibility to  focus the beam ({\em beam focusing}) at a specific location, in contrast to only a specific direction as in far-field conditions via conventional {\em beam  steering}. Beam focusing gives rise to the possibility to support multiple coexisting orthogonal links,  even at similar angles
\cite{nepa2017near}.   


Most existing works on near-field focusing appeared in the antenna theory literature (see, e.g., \cite{buffi2012design,liu2019antenna} and the references therein),  wherein the \ac{em} field in the Fresnel region was characterized and modeled for various antenna implementation technologies. For example, the authors in \cite{buffi2012design} studied the effect of the planar array's antenna size, inter-element distance, and focal distance on the near-field focusing performance. In \cite{liu2019antenna}, the authors developed a multi-focus antenna array to focus signals on multiple focal points in the near-field region. While antenna theory provides tools to achieve beam focusing, how to exploit this ability to facilitate near-field wireless communications\footnote{Note that in this paper, with ``near-field'' we refer only to the Fresnel region (also known as the radiative near-field), thus neglecting the reactive near-field region, that entails distances in the order of the wavelength.} is still in its infancy, and only a small set of works have studied near-field focusing from a communication perspective.  
In \cite{nishimori2011transmission}, the authors considered a point-to-point short-range \ac{mimo} communication system, which consists of two identical transceiver array antennas that face each other with a distance comparable to the size of the antenna aperture. More recently, near-field communications with antennas based on \acp{lis}, whose large aperture gives rise to near-field operation, was explored in \cite{Dar:2020,torres2020near,tang2020wireless,garcia2020reconfigurable}. In particular, \cite{Dar:2020} considered a single-user scenario, and characterized the path-loss and the available communication modes, while  \cite{torres2020near} studied a two-user uplink scenario in which the \ac{bs} is equipped with an \ac{lis}. Both \cite{Dar:2020} and \cite{torres2020near} studied ideal antenna architectures, where the transceiver has direct access to the signal observed at each element.
 In addition, \cite{tang2020wireless, garcia2020reconfigurable} considered near-field communication in  \acp{ris}-assisted systems, where \acp{ris} act as a passive reflector. Nonetheless, the potential of near-field focusing in facilitating  massive \ac{mimo} downlink communications with practical antenna technologies has not been thoroughly studied to date.


The ability to achieve focused beams in massive \ac{mimo} systems is highly dependent on the signal processing capabilities of the antenna array, which  vary between different architectures. The most flexible solution for a given array of radiating elements is the fully-digital architecture, where  each antenna element is connected to a dedicated radio frequency (RF) chain. In such architectures, the transceiver is capable of controlling beams at infinitely many directions at the same time, which greatly enhances the spatial flexibility \cite{zhang2005variable}. However, towards the deployment of large-scale arrays in 5G and beyond communication systems, the implementation of a fully-digital architecture becomes extremely challenging due to its increased cost and power consumption. To alleviate this, hybrid analog/digital architectures 
are commonly considered for massive \ac{mimo} communications \cite{sohrabi2016hybrid,ioushua2019family,mendez2016hybrid}. Such hybrid architectures operate with fewer RF chains than antenna elements by combining low-dimensional digital processing and high-dimensional analog precoding, typically implemented using an interconnection of phase-shifters. An alternative emerging technology for efficiently realizing large-scale arrays is based on \acp{dma}. \acp{dma} are a practical implementation of \acp{lis}, i.e., they  enable programmable control of the transmit/receive beam patterns, which also provide advanced analog signal processing capabilities  \cite{shlezinger2019dynamic,DSmith-2018TCOM,wang2019dynamic,wang2019dynamic2,shlezinger2020dynamic}, and  naturally implement RF chain reduction without  dedicated analog circuitry. 
Furthermore, \acp{dma} facilitate densification of the antenna elements which can be exploited to improve the focusing performance.
The fact that the signal processing capabilities of the aforementioned antenna architectures affect their ability to generate focused beams motivates the study of near-field multi-user communications with these different antennas. 

In this paper we study multi-user downlink MIMO systems operating in the near-field region. We focus on the exploration of utilizing various antenna architectures, including fully-digital arrays, phase-shifters based hybrid architectures, and \acp{dma}, to facilitate multi-user communications via near-field signalling. In particular, we aim to quantify the capabilities of massive \ac{mimo} architectures in forming focused beams, as well as the effect of such an operation on downlink multi-user systems.  To the best of our knowledge, this work represents the first study on the design of focused beams (e.g., beam focusing) as means of optimizing multi-user communication objectives, and utilizing this capability to facilitate simultaneous communication with multiple users.

We begin by formulating a mathematical model for downlink near-field multi-user MIMO systems. Our model
incorporates both the digital signal and analog signal processing carried out  by the \ac{bs},  as well as the propagation of the transmitted EM waves in near-field wireless communications. Then, we study beam focusing design in order to maximize the sum-rate under each of the considered antenna architectures; We first consider fully-digital antenna systems, which is the most flexible architecture as it allows independent control of  the signal fed to each transmitting element. We then use the obtained fully-digital beam focusing configuration as a baseline for deriving the corresponding setting for phase shifter based  hybrid architectures. 
%
In particular, we show that the sum-rate maximization problem for such hybrid antennas can be tackled using manifold optimization techniques for sum-rate optimization in far-field communications aided by \acp{ris} \cite{yu2019miso,guo2020weighted, pan2020multicell,AbrDarDiR:21}.  
For \acp{dma}, where the analog signal processing capabilities follow the Lorentzian-form response of metamaterial elements \cite{Smith-2012APM}, we cannot adapt design methods previously proposed for far-field systems as we do for phase shifter based hybrid antennas, and thus we derive a dedicated configuration algorithm. To that aim, we first focus on a single-user case, for which we are able to optimize the \ac{dma} configuration.  Then, we consider the case of multiple users, where the resulting  optimization problem is non-convex, and propose an alternating design algorithm to jointly optimize the \ac{dma} configuration and digital precoding.

While parts of our technical derivations build upon design algorithms proposed for far-field communications, we demonstrate that the incorporation of the near-field characteristics results in fundamentally different beam patterns compared to the far-field. In particular, simulation results show that our proposed designs for different types of antenna architectures are all capable of concentrating the transmissions to the desired focal points, illustrating the beam focusing ability of our proposed designs. Furthermore, it is demonstrated that by exploiting the beam focusing capabilities of near field transmissions via the proposed configuration methods, one can reliably simultaneously communicate with multiple users located in the same angular direction with different ranges, which is not achievable using conventional beam steering techniques. Finally, we show that by accounting for the near-field capabilities in transmission, one can achieve notable gains in  achievable sum-rate compared to  designs assuming conventional far-field operation.

The rest of this paper is organized as follows: Section \ref{sec:System_model} presents the near-field channel model, reviews the considered antenna architectures, and formulates the near-field-aware precoding problem. Section \ref{sec:Solution} 
presents efficient algorithms for tuning the beam focus solutions for all considered antenna architectures, while Section \ref{sec:Sims} numerically demonstrates the beam-focusing ability and evaluates its effect on the achievable  rates. Finally, Section \ref{sec:Conclusions} concludes the paper.

Throughout the paper, we use boldface lower-case and upper-case letters for vectors and matrices, respectively. Calligraphic letters are used for sets. The $\ell_2$ norm, vectorization operator,  transpose, conjugation, Hermitian transpose, trace, Kronecker product, Hadamard product, and stochastic expectation are written as $\| \cdot \|$,  ${\rm Vec}(\cdot)$, $(\cdot)^T$, $(\cdot)^{\dag}$,  $(\cdot)^H$, ${\rm {Tr}}\left(\cdot\right)$, $\otimes$, $\circ$, and $\E\{ \cdot \}$,  respectively.  Finally,  for any vector ${\bf x}$, $({\bf x})_{i}$ denotes the $i$th entry of ${\bf x}$.

	
	\vspace{-0.2cm}
	\section{System Model}
	\label{sec:System_model}
	\vspace{-0.1cm}

 In this section, we describe the considered near-field multi-user MIMO communication system.  We first introduce the concept of near-field transmission in Section \ref{sec:Near-field region}. Then, we formulate the mathematical model of near-field wireless channels in Section \ref{sec:model}. After that, we present three types of antenna architectures and their corresponding signal models in Section \ref{sec:antenna}. Finally, in Section  \ref{sec:problem_formulation},  we  formulate the optimization problem of designing the transmission beam pattern to maximize the sum-rate for near-field communications.

	\vspace{-0.2cm}
\subsection{Near-Field Region}
	\label{sec:Near-field region}
	\vspace{-0.1cm}
  
According to conventional notation, transmission is considered to take place in the far-field if the distance between the transmitter and the receiver is larger than the Fraunhofer distance, denoted by $d_\mathrm{F}=\frac{2\,D^2}{\lambda}$, where $D$ is the antenna diameter and $\lambda$ the wavelength. For distances larger than $d_F$, the signal wavefront can be faithfully approximated as being planar. When the distance is shorter than $d_\mathrm{F}$ but larger than the Fresnel distance, typically denoted by $d_\mathrm{N}=\sqrt[3]{\frac{D^4}{8\,\lambda}}$, the receiver is considered to lie in the radiative near-field Fresnel region, referred to henceforth as the near-field region. The near-field accounts for distances in between $d_\mathrm{F}$ and $d_\mathrm{N}$. 
%
The boundary $d_\mathrm{N}$ constitutes the minimal distance from which reactive field components from the antenna itself can be neglected.
\begin{figure} 
  \centering 
  \subfigure[$d_\mathrm{F}$ vs. antenna diameter.]{ 
  \centering
    \label{fig:subfig:Fraunhofer}
    \includegraphics[width=0.45\columnwidth]{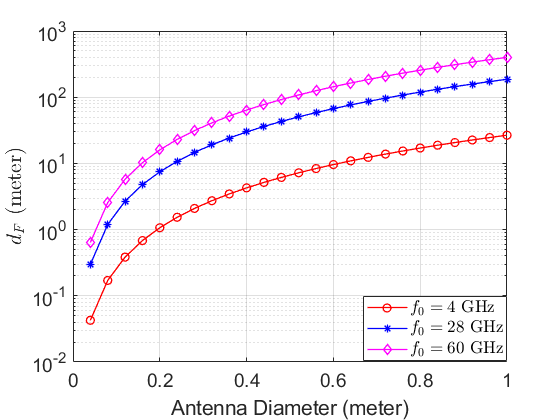} 
  } 
  \quad
  \subfigure[$d_\mathrm{N}$ vs. antenna diameter.]{ 
  \centering
    \label{fig:subfig:Fresnel_d} 
    \includegraphics[width=0.45\columnwidth]{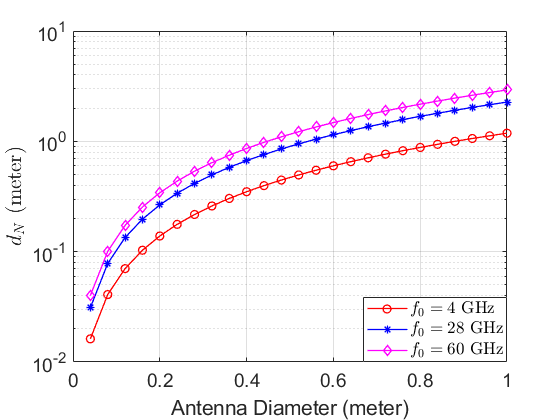}
    } 
  \caption{Near-field region boundaries for different frequencies and antenna diameter values.} 
  \label{fig:Fresnel_region} 
\end{figure}

Conventional wireless communications lie in the far-field due to the entailed distances, antenna sizes and frequencies. For instance, for an antenna of diameter $D= 0.1$ meters at  carrier frequency of $5$ GHz, any receiver located at a distance of more than $d_F=0.33$ meters is considered to lie in the far-field. 
However, for mmWave frequency bands, particularly when combined with antenna arrays of relatively large physical size,  this approximation no longer holds, and one must account for the spherical wavefront shape. This is demonstrated in Fig.~\ref{fig:Fresnel_region}, which illustrates the values of $d_\mathrm{F}$ (upper limit) and $d_\mathrm{N}$ (lower limit) for different antenna diameters $D$ and carrier frequencies $f_0$, that together delimit the expected near-field operating region.    
From Fig.~\ref{fig:Fresnel_region}, we can clearly see that when the system operates at mmWave frequency bands, the near-field distance can be up to dozen of meters for relative small antennas/surfaces. For instance, for a \ac{bs} equipped with an antenna of diameter $D= 0.5$ meters at  carrier frequency $28$ GHz, any user closer than $47$ meters from the antenna resides in its near-field.
Therefore, it is of interest to investigate how to exploit the non-negligible spherical wavefronts of the near-field  to increase communication rates. In particular, this gives rise to the possibility of generating focused beams and to enhance the communication performance of wireless networks by alleviating multi-user interference.

\vspace{-0.2cm}	
\subsection{Near-Field Channel Model} \label{sec:model}
\begin{figure}
		\centering	
	\includegraphics[width=0.8\columnwidth]{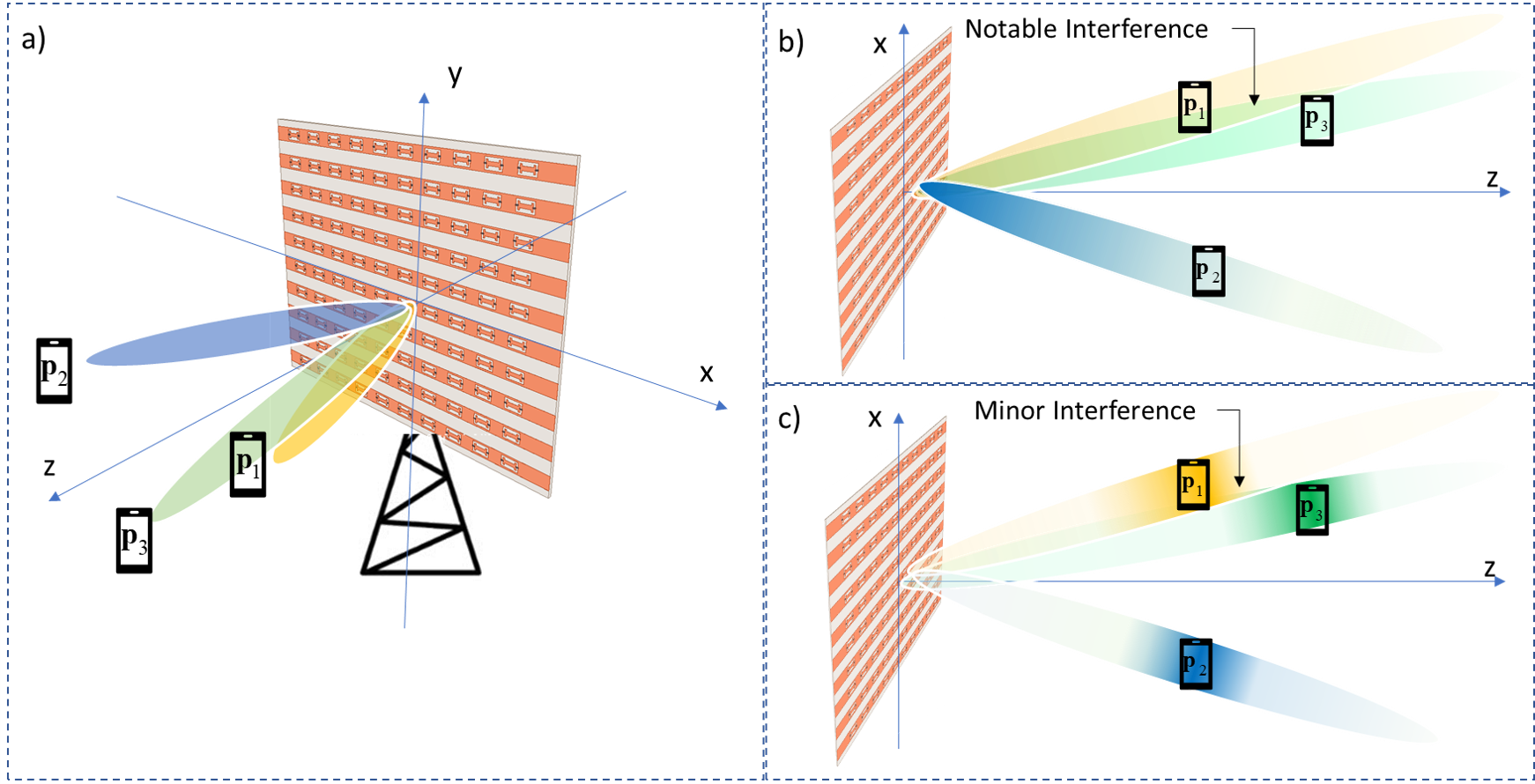}
		\vspace{-0.4cm}
		\caption{Near-field communications with $M=3$ receivers, with dedicated beams directed towards each user: $(a)$ illustration in three-dimensional space; $(b)$ beam steering based on far-field design, resulting in notable interference among users sharing the same angular direction; $(c)$ beam focusing, resulting in minor interference.} 
		\label{fig:system_model}
	\end{figure}

\vspace{-0.1cm}	
To evaluate the ability to exploit near-field operation in \ac{mimo} communications, we focus on downlink multi-user systems. In particular, we consider a downlink multi-user MIMO system  where the \ac{bs}  employs a \ac{upa}, i.e., a two-dimensional antenna surface, with $N_e$ uniformly spaced radiating elements in the horizontal direction and $N_d$ elements in the  vertical direction. The total number of antenna elements is thus $N=N_d\, \times N_e$. 
We denote the Cartesian coordinate of the $l$th element of the $i$th row as $\pil=(x_l,y_i,0)$, $l=1,2, \ldots N_e$, $i=1,2, \ldots N_d$.
The BS communicates with $M$ single-antenna receivers, as illustrated in Fig. \ref{fig:system_model}. We consider that the receivers' positioning information is known at the BS via high-accuracy wireless positioning techniques\cite{guidi2019radio}. We focus on communications in the near-field, i.e., where the distance between the \ac{bs} and the users is not larger than the Fraunhofer distance $ \dF$ and not smaller than the Fresnel limit $d_{\mathrm N}$. The properties of near-field spherical waves allow for the generation of focused beams to facilitate communications.

 To start, we model the near-field wireless channels.  The signal received in free-space conditions by the $m$th  user, $m \in \mySet{M} \triangleq \{1,2,\ldots,M\}$, located at $\mathbf{p}_m=(x_m,y_m,z_m)$ is given by 
\begin{equation} \label{eq:near-field_wireless_channel}
  r(\mathbf{p}_m) 
 	 = \sum_{i=1}^{N_d} \sum_{l=1}^{N_e} \Ail (\p_m)\, e^{ -\jmath k |\p_m-\pil|} \,s_{i,l}\, +n_m, 
 \end{equation} 
where  $s_{i,l}$ denotes the signal emitted by the  antenna at position $\pil$; 
the term $e^{ -\jmath k |\p_m-\pil|}$ contains the phase due to the distance travelled by the wave from $\pil$ to $\p_m$;  $k=2\pi /\lambda$ is the wave number;   $\Ail(\p_m)$ denotes the channel gain coefficient; and  $n_m \sim \mathcal{C} \mathcal{N}\left(0, \sigma^{2}\right)$ is the \ac{awgn} at user $m$. Following \cite{ellingson2019path}, we write  
\begin{align} 
\label{eqn:Adef}
\Ail(\p_m)=\sqrt{F(\Theta_{i,l,m})}\frac{\lambda}{4\,\pi |\p_m-\pil|} \, ,
\end{align}
where $\Theta_{i,l,m}=(\theta_{i,l,m},\phi_{i,l,m})$ is the elevation-azimuth pair from the $l$th element of the $i$th row to the $m$th user, while $ F(\Theta_{i,l,m})$ is the radiation profile of each element, modeled as 
\begin{align}
    F(\Theta_{i,l,m}) \!=\! \left\{\begin{array}{ll} 2(b+1) \cos^b (\theta_{i,l,m})  & \,  \theta_{i,l,m} \in [0,\pi/2] \, ,   \\0 & \, \text{otherwise}.  \\ \end{array}\right.    
    \label{eqn:Boresight}
 \end{align}
 %
 In \eqref{eqn:Boresight}, the parameter $b$  determines the Boresight gain, whose value depends on the specific technology adopted  \cite{ellingson2019path}. As an example, for the dipole case we have $b=2$, which yields $F(\Theta_{i,l,m})=6\,\cos^2 \theta_{i,l,m}$. Here, the model accounts for the fact that the transmitted power is doubled by the reflective ground behind the antenna.  
 %
%



To obtain a more compact formulation of  the received signal  in \eqref{eq:near-field_wireless_channel}, we define the vector
\ifsingle
\begin{equation} 
{\bf a}_m(\mathbf{p}_m)  =\big[A_{1,1}(\p_m)\, e^{ -\jmath k |\p_m-\p_{1,1}|},A_{1,2}(\p_m)\, e^{ -\jmath k |\p_m-\p_{1,2}|}, \cdots,A_{N_d,N_e}(\p_m) e^{ -\jmath k |\p_m-\p_{N_d,N_e}|}\big]^H.
\label{eqn:AmDef}
\end{equation}
\else
\begin{equation} 
\begin{split}
{\bf a}_m &=\big[A_{1,1}(\p_m)\, e^{ -\jmath k |\p_m-\p_{1,1}|},A_{1,2}(\p_m)\, e^{ -\jmath k |\p_m-\p_{1,2}|}, \\
&~~~~~\cdots,A_{N_d,N_e}(\p_m) e^{ -\jmath k |\p_m-\p_{N_d,N_e}|}\big]^H.
\end{split}
\label{eqn:AmDef}
\end{equation}
\fi

For convenience, we omit the location index $\mathbf{p}_m$ in ${\bf a}_m(\mathbf{p}_m)$ for the rest of this paper.  Using \eqref{eqn:AmDef}, we can then write the received signal at the $m$th user as
\begin{equation} \label{eq:RX2_vector_compact}
r(\mathbf{p}_m) ={\bf a}_m^H {\bf s} +n_m, \quad  m\in\mySet{M},
\end{equation}
where ${\bf s}=\left[s_{1,1},s_{1,2}\cdots ,s_{N_d,N_e}\right]$ collects the transmitted signals of all antennas.

Near-field operation is encapsulated in the vector ${\bf a}_m$ defined in \eqref{eqn:AmDef}. When the far-field approximation holds, the outputs of all the elements experience the same path loss (e.g., $A(\p_m)=A_{i,l}(\p_m)\,\, \forall \,i,l$) and a phase shift with constant gradient along the array aperture, with $\Theta_{i,l,m}=\Theta_{m}\,\forall\,i,l$. More specifically, ${\bf a}_m$ becomes the traditional beamsteering vector given by $    {\bf a}_m=A(\p_m)\,\left[e^{-jk \Psi_{1,1}(\Theta_m)}, \ldots, e^{-jk \Psi_{i,l}(\Theta_m)}, \ldots, e^{-jk \Psi_{N_d,N_e}(\Theta_m)}  \right]$, where $\Psi_{i,1}(\Theta_m)$ depends only on the direction of the $m$th user and on the spacing among the radiating elements.
The diversity among the elements of ${\bf a}_m$ in the near-field gives rise to the possibility to focus the beam towards an intended position in space, rather than just steer it at a given angle, as enabled in the far-field. 




\vspace{-0.2cm}
\subsection{Antenna Architectures}
\label{sec:antenna}
\vspace{-0.1cm}

The beam pattern in \eqref{eq:RX2_vector_compact} depends on the transmitted signal $\myVec{s}$, which in turn depends on the signal processing capabilities supported by the antenna architecture. We consider three types of antenna schemes as shown in Fig. \ref{fig:antenna archtecture}: a fully-digital architecture, phase-shifters based analog precoder, and a \ac{dma}. In the following subsections, we will introduce each of the antenna architectures and provide insights on the corresponding signal model of $\myVec{s}$.

\subsubsection{Fully-digital antenna}

\begin{figure} 
  \centering 
  \subfigure[The fully-digital architecture]{ 
    \label{fig:subfig:fully_digital}
    \includegraphics[width=2.8in]{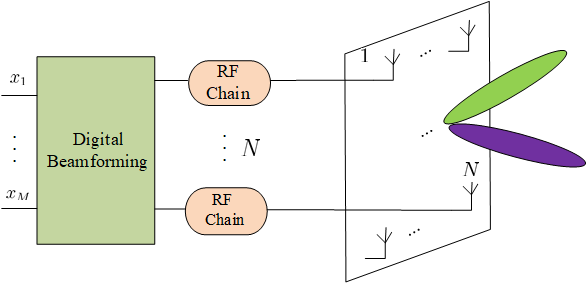} 
  } 
  \subfigure[The phase-shifters based hybrid architecture]{ 
    \label{fig:subfig:hybrid} 
    \includegraphics[width=3.0in]{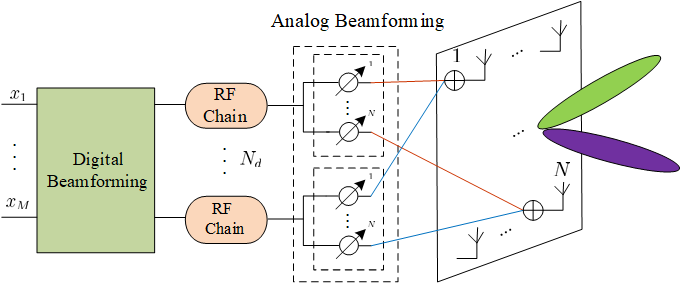}
    } 
  \subfigure[DMA-based architecture.]{ 
    \label{fig:subfig:DMA}
    \includegraphics[width=2.8in]{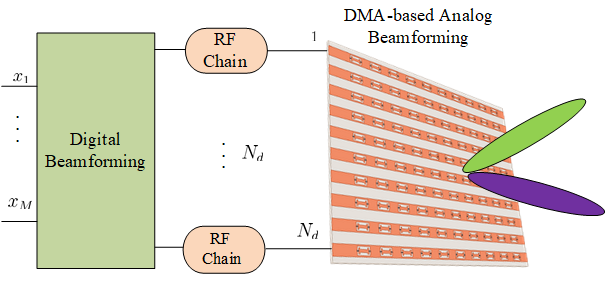} 
  } 
  \caption{Three typical antenna architectures.} 
  \label{fig:antenna archtecture} 
\end{figure}

In fully-digital antennas each element is connected to a dedicated RF chain, as illustrated in Fig.~\ref{fig:subfig:fully_digital}. Such architectures provide the most flexible signal processing capabilities, as the input to each element can be separately processed. Nonetheless, fully-digital antennas are typically costly, particularly in massive \ac{mimo} systems, since the number of RF chains is equal to the number of antenna elements \cite{ioushua2019family,mendez2016hybrid}. Thus, we consider the fully-digital architecture as a baseline system, representing the beam focusing capabilities achievable in near-field multi-user communications with unconstrained linear precoding. In this case, the signal transmitted by the \ac{bs} can be written as
\begin{equation} \label{eq: base_band_fully}
{\bf s} =\sum_{m=1}^M \tilde{\bf w}_m x_m,
\end{equation}
where $x_m$ is the normalized data symbol intended for the $m$th user, i.e., $\mathbb{E}[x_m^2]=1$,  and $\tilde{\bf w}_m \in {\mathbb{C}}^{N }$ is the  precoding vector for $x_m$.

By expressing the channel input $\mathbf{s}$ via \eqref{eq: base_band_fully}, the received signal of the $m$th user is given by
\begin{equation} \label{eqn:RX2_fully}
r(\mathbf{p}_m) ={\bf a}_m^H\,\sum_{j=1}^M \tilde{\bf w}_j x_j +n_m, \quad  m\in\mySet{M}.
\end{equation}
Based on \eqref{eqn:RX2_fully}, the achievable rate of the $m$th user for the fully-digital antenna case is given by
\begin{equation} \label{eqn:sum-rate_digital}
R_m\left(\{\tilde{\bf w}_m\} \right) =  {\rm log}_{2}\left(1+\frac{\left|{\bf a}_m^H\, \tilde{\bf w}_m \right|^{2}}{\sum_{j \neq m}\left|{\bf a}_m^H\, \tilde{\bf w}_j \right|^{2}+\sigma^{2}}\right), \quad m \in \mySet{M}.
\end{equation}
Expression \eqref{eqn:sum-rate_digital} characterizes the achievable rate at each user for a given precoding configuration, and is computed assuming that the users treat the interference as noise. While in some scenarios one can achieve higher rates by decoding the interference \cite{shlezinger2018spectral}, the common practice in downlink massive \ac{mimo} systems is to treat it as noise \cite{marzetta2010noncooperative}, resulting in the rate in \eqref{eqn:sum-rate_digital}.

\subsubsection{Phase shifter based hybrid antenna}
Hybrid antennas combine digital signal processing with some constrained level of analog signal processing. Here, the number of RF chains, denoted by $N_{\rm RF}$, is smaller than the number of antenna elements $N$. In fully-connected phase shifting analog precoders, each RF chain output is connected to all the transmit antennas through a phase-shifters based analog beamforming network, as shown in Fig. \ref{fig:subfig:hybrid}. 
In this case, the signal transmitted by the BS is given by
\begin{equation} \label{eq: base_band_hybrid}
{\bf s} = \sum_{m=1}^M {\bf Q}\, {\bf w}_m x_m.
\end{equation}
Here, the digital precoding vector ${\bf w}_m$ is an $N_{\rm RF} \times 1$ vector, while  $\myMat{Q} \in {\mathbb{C}}^{N \times N_{\rm RF}}$ represents analog precoding, which maps the $N_{\rm RF}\times 1$ digital vector into the $N$ antenna elements. For phase-shifters based analog precoding, the elements of $\myMat{Q}$, denoted by $\{q_{i,l}\}$, satisfy 
\begin{equation}\label{eqn:analog_constraint}
q_{i,l}\in \mathcal{F}\triangleq\left\{e^{j \phi}| \phi \in [0,2\pi]\right\},~ \forall i,l.
\end{equation}
%
Comparing \eqref{eq: base_band_hybrid} with \eqref{eq: base_band_fully}, it holds that the received signal model for the hybrid precoder is a special case of the fully-digital one, obtained by setting $\tilde{\bf w}_m = \myMat{Q}\myVec{w}_m$ for each $m\in\mathcal{M}$. As a result, the achievable rate for a given configuration of the analog precoder $\myMat{Q}$ and digital vectors $\{\myVec{w}_j\}$ is given by $R_m(\{\myMat{Q}\myVec{w}_j\})$, computed via \eqref{eqn:sum-rate_digital}.


\subsubsection{DMA}


\acp{dma} utilize radiating metamaterial elements embedded onto the surface of a waveguide to realize reconfigurable antennas of low cost and power consumption \cite{shlezinger2020dynamic}. The typical \ac{dma} architecture is comprised of multiple waveguides, e.g. microstrips, and each microstrip contains multiple metamaterial elements. The elements are typically sub-wavelength spaced, implying that one can pack a larger number of elements in a given aperture compared to conventional architectures based on, e.g., patch arrays \cite{shlezinger2020dynamic}. The frequency response of each individual element can be externally adjusted by varying its local electrical properties   \cite{Sleasman-2016JAWPL}.

\begin{figure}
		\centering	
	\includegraphics[width=6in]{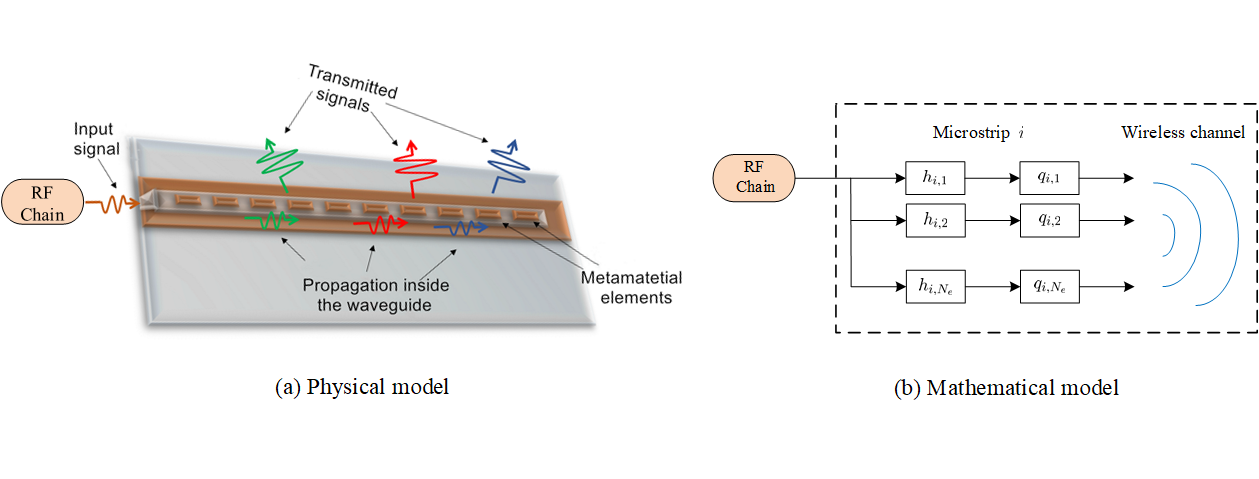}
		\vspace{-1.5cm}
		\caption{Illustration of signal transmission using a microstrip. (a) Physical model; (b) Mathematical model.} 
		\label{fig:Microstrip_illustration}
	\end{figure}

For DMA-based transmitting architectures, each microstrip is fed by one RF-chain, and the input signal is  radiated by all the elements located on the microstrip, as shown in Fig. \ref{fig:subfig:DMA}. Fig.~\ref{fig:Microstrip_illustration} shows an example of transmitting a signal using a single microstrip with multiple elements. To formulate its input-output relationship, consider a  DMA consisting of $N=N_d \cdot N_e$  metamaterial elements, where here $N_d$ and $N_e$ are the numbers of microstrips and elements in each microstrip, respectively. The equivalent baseband signal radiated from the $l$th  element of the $i$th microstrip  is
	%
$s_{i,l} = h_{i,l} \, q_{i,l}\, z_i  $,
where $z_i$ is the baseband signal fed to the $i$th microstrip, $q_{i,l}$ denotes the tunable  response of the $l$th element of the $i$th microstrip, and $h_{i,l}$ encapsulates the effect of signal propagation inside the microstrip. We consider the case where the response of the  elements is frequency flat as in \cite{shlezinger2019dynamic}, and focus on the Lorentzian-constrained phase model of the metamaterial elements frequency response 
\cite{DSmith-2017PRA, smith2017analysis}, i.e., 
\begin{equation}\label{eqn:FreqSel}
q_{i,l} \in \mathcal{Q}\triangleq\left\{\frac{j+e^{j \phi}}{2}| \phi \in [0,2\pi]\right\},~ \forall i,l.
\end{equation}
The   signal propagation inside the microstrip is formulated as
\begin{equation} \label{eq:delay_in_microstrip}
    h_{i,l}= e^{-\rho_{i,l}(\alpha_i+ j\beta_i) },~ \forall i,l
\end{equation}
where $\alpha_i$ is the waveguide attenuation coefficient, $\beta_i$ is the wavenumber, and $\rho_{i,l}$ denotes the location of the $l$th element in the $i$th microstrip. 

Letting ${\bf z} =[z_1,\ldots,z_{N_d}]^T$ be the microstrips input, the baseband representation of the signal transmitted by the DMA output is given by
%
${\mathbf{s}}= {\mathbf{H} \,\mathbf{Q}\, \mathbf{z}}$,
%
where 
$\mathbf{H}$ is a $N \times N$ diagonal matrix with elements ${\mathbf{H}}_{((i-1)N_e+l,(i-1)N_e+l)}=h_{i,l}$, and $\mathbf{Q}\in {\mathbb{C}}^{N \times N_d}$ denotes the configurable weights of the DMAs, with each element given by
\begin{equation} \label{eq: weighting_matrix}
{\mathbf{Q}}_{(i-1) N_{e}+l, n}=\left\{\begin{array}{ll}
q_{i, l} & i=n, \\
0 & i \neq n.
\end{array}\right.
\end{equation}
%
The DMA input signal is given by ${\bf z} =\sum_{m=1}^M {\bf w}_m x_m $, where  ${\bf w}_m \in {\mathbb{C}}^{N_d }$ is the digital precoding vector for $x_m$. The baseband channel input transmitted by the \ac{dma} is thus given by
%
$\mathbf{s}=\sum_{m=1}^M  \mathbf{H} \mathbf{Q} {\bf w}_m x_m \,$.
%
We again note that the transmitted signal is formally equal to that of a fully-digital architecture with precoding vectors $\tilde{\myVec{w}}_m = \myMat{H}\myMat{Q}\myVec{w}_m$ for each $m \in \mathcal{M}$. Consequently, the resulting achievable rate of the $m$th user for a given \ac{dma} configuration matrix $\myMat{Q}$ and digital precoding vectors $\{\myVec{w}_j\}$ is computed as $R_m(\{\myMat{H}\myMat{Q}\myVec{w}_j\})$ using  \eqref{eqn:sum-rate_digital}.

\vspace{-0.2cm}
\subsection{Problem Formulation}
\label{sec:problem_formulation}
\vspace{-0.1cm}

Based on the above model,  we investigate multi-user communications in the near-field, considering the possibility of achieving reliable communications when different users share  similar directions but are located at different distances from the \ac{bs}. The aim here to is to design the transmission beam pattern to maximize the achievable sum-rate, reflecting the overall number of bits which can be reliably conveyed per channel use. Based on the different antenna architectures, for a given transmit power constraint $P_{\rm max}>0$, the task of interest can be written as:
\begin{align} \label{eq:Original_optimization}
&\max_{ \left\{\tilde{\bf w}_m\right\} }~~\sum_{m=1}^{M} R_m\left(\{\tilde{\bf w}_j\}_{j\in\mathcal{M}}\right) = \sum_{m=1}^{M} {\rm log}_{2}\left(1+\frac{\left|{\bf a}_m^H\, \tilde{\bf w}_m \right|^{2}}{\sum_{j \neq m}\left|{\bf a}_m^H\, \tilde{\bf w}_j \right|^{2}+\sigma^{2}}\right), \\
&~~~~s.t.~~~~~  \sum_{m=1}^{M} \left\|\tilde{\bf w}_m\right\|^2 \leq P_{\rm max}, \quad  \left\{\tilde{\bf w}_m\right\} \in \mySet{W}. \notag
\end{align} 
%
%
%
%
%
%
The problem formulated in \eqref{eq:Original_optimization} is similar to the corresponding problem encountered in far-field communications. The fact that communications is carried out in the near-field is encapsulated in the vectors $\{\myVec{a}_m\}$. As a result, some of the tools used for tackling this problem in the sequel are adopted from studies considering far-field communications. 

The set of feasible precoding vectors $\mySet{W}$ in \eqref{eq:Original_optimization} captures the unique constraints imposed by the antenna architecture: for fully-digital \acp{upa}, $\mySet{W}_{\rm FD}$ is the set of all $M$-tuples of vectors in $\mathbb{C}^{N}$; For hybrid beamformers, the feasible set $\mySet{W}_{\rm HB}$ is expressed as
\begin{equation}
\label{eqn:FeasibleHBF}
    \mySet{W}_{\rm HB} = \{\{\tilde{\myVec{w}}_m\}_{m\in\mathcal{M}} | \tilde{\myVec{w}}_m = \myMat{Q}\myVec{w}_m; \myMat{Q}\in\mySet{F}^{N\times N_d}\};
\end{equation}
For \acp{dma}, the set of feasible precoders can be written as 
\begin{equation}
\label{eqn:FeasibleDMA}
    \mySet{W}_{\rm DMA} = \{\{\tilde{\myVec{w}}_m\}_{m\in\mathcal{M}} | \tilde{\myVec{w}}_m = \myMat{H}\myMat{Q}\myVec{w}_m\},
\end{equation}
where $\myMat{H}$ is the fixed diagonal matrix representing the propagation inside the microstrips, while $\myMat{Q}$ can be written as in \eqref{eq: weighting_matrix} in which the non-zero elements take values in $\mySet{Q}$ defined in \eqref{eqn:FreqSel}.

For both the hybrid antenna and the DMA architectures, we optimize the corresponding $\left\{{\bf w}_m\right\}$ and ${\bf Q}$ to obtain the feasible precoders $\left\{\tilde{\myVec{w}}_m\right\}$. For the hybrid antenna, each element of ${\bf Q}$ need to satisfy the unit modulus constraint in \eqref{eqn:analog_constraint}, whereas for the DMA, the non-zero elements of ${\bf Q}$ should satisfy the  Lorentzian constraint in \eqref{eqn:FreqSel}.

While the formulation of  \eqref{eq:Original_optimization} does not explicitly account for the fact that transmission takes place in the near-field, this property is embedded in the equivalent channel vectors $\{\myVec{a}_m\}$. As we show in the sequel, while the objective in \eqref{eq:Original_optimization} appears invariant of the shape of the resulting beams, maximizing the achievable sum-rate in the near field yields focused beams allowing to mitigate interference between users lying in the same angular direction. 



\vspace{-0.2cm}
\section{Beam Focusing Aware Precoding Design}
\label{sec:Solution}
\vspace{-0.1cm}

In this section, we study beam focusing-aware precoding  design to maximize the sum-rate.  We begin with unconstrained fully-digital antennas in Section \ref{sec:fully_digital}. Then, we derive the hybrid phase shifter setting  in Section \ref{sec:hybrid}, and consider DMA-based antenna architectures in Section \ref{sec:DMA}. We conclude with a discussion in Section~\ref{sec:Discussion}.

\vspace{-0.2cm}
\subsection{Fully-Digital Beam Focusing}\label{sec:fully_digital}	
\vspace{-0.1cm}

For the fully-digital beam focusing design, the feasible precoding set $\mySet{W}_{\rm FD}$ is unconstrained, and includes all combinations of $M$ vectors in $\mathbb{C}^N$. 
%
For the single-user case, i.e., $M=1$, the rate in \eqref{eq:Original_optimization} is maximized by setting $\tilde{\bf w}_1 = \sqrt{P_{\rm max}}\frac{{\bf a}_1}{\left|{\bf a}_1\right|}$. However, for the general case of $M>1$, problem $\eqref{eq:Original_optimization}$ is non-convex,  and thus it is difficult to find the optimal solution. Nevertheless, due to the similarity between \eqref{eq:Original_optimization}  and the corresponding sum-rate maximization for interference broadcast channels operating in the far-field, one can utilize tools derived for far-field systems. A candidate strategy is to use the weighted sum \ac{mse} minimization approach \cite{shi2011iteratively} to deal with problem $\eqref{eq:Original_optimization}$,   which  guarantees convergence to a stationary point.

By exploiting the relationship between sum-rate maximization and \ac{mse} minimization \cite[Thm. 1]{shi2011iteratively}, we have the following lemma.

\begin{lemma}\label{Lemma1}
Problem \eqref{eq:Original_optimization} with $\mySet{W}=\mySet{W}_{\rm FD}$ is equivalent (in the sense of having the same global optimum) to the following problem
\begin{equation} \label{eq:problem_digital_w}
\begin{split}
&\max_{ \left\{\tilde{\bf w}_m, u_m, v_m\right\}}~~~\sum_{m=1}^{M}~{\rm log}_2 (v_m) - v_m e_m \left(u_m,\left\{\tilde{\bf w}_m\right\}\right)\\
&~~~~~~~s.t.~~~~~~~~\sum_{m=1}^{M} \left\|\tilde{\bf w}_m\right\|^2 \leq P_{\rm max}, \quad v_m \geq 0,~ m\in \mySet{M},
\end{split}
\end{equation} 
where $u_m$ and $v_m$ are auxiliary variables, and $e_m\left(u_m,\left\{\tilde{\bf w}_m\right\}\right)$ is given by
$e_m\left(u_m,\left\{\tilde{\bf w}_m\right\}\right) = \left|1-u_m\, {\bf a}_m^H\, \tilde{\bf w}_m \right|^2  
+   \sum_{j \neq m}\left|u_m\,{\bf a}_m^H\,  \tilde{\bf w}_j\right|^{2}  + \sigma^2\,\left|u_m\right|^2$. 
\end{lemma}



Although problem \eqref{eq:problem_digital_w} involves more optimization variables than \eqref{eq:Original_optimization}, it is  concave for each set of the optimization variables when
the remaining two sets are fixed. Hence, the block coordinate descent method can be applied to solve  \eqref{eq:problem_digital_w}, resulting in the procedure summarized as Algorithm~\ref{algorithm_digital}, which is based on the method proposed in \cite[Sec. III]{shi2011iteratively}. 
 \begin{algorithm}
 \caption{Alternating optimization of fully-digital precoders}
 \label{algorithm_digital}
 \begin{algorithmic}[1]
 \renewcommand{\algorithmicrequire}{\textbf{Initialize:}} \REQUIRE 
 $\left\{\tilde{\bf w}_m^0\right\}_{m=1}^M$; \\
\FOR{$t=1,\ldots,t_{\rm max}$} 
  \STATE Update $u_m^{t} = \frac{{\bf a}_m^H\, \tilde{\bf w}_m^{t-1}}{\sum_{j=1}^M \left|{\bf a}_m^H\,\tilde{\bf w}_j^{t-1}\right|^{2} + \sigma^2},~ \forall m$;
  \STATE Update $v_m^t = (e_m\left(u_m^t,\left\{\tilde{\bf w}_m^{t-1}\right\}\right))^{-1},~\forall m$;
  \STATE \label{stp:update1} Update $\tilde{\bf w}_m^t \!=\! u_m^t\,v_m^t\,\left( \sum_{j=1}^M\,v_j^t\,\left|u_j^t\right|^2{\bf a}_j{\bf a}_j^H \!+\! \lambda_p {\bf I} \right)^{-1}{\bf a}_m, \forall m$;  \\
 \ENDFOR
 \renewcommand{\algorithmicrequire}{\textbf{Output:}} \REQUIRE  $\{\myVec{w}_m^t\}$.
 \end{algorithmic} 
 \end{algorithm}

In Algorithm \ref{algorithm_digital}, the parameter $\lambda_p$ in step \ref{stp:update1} is the Lagrangian multiplier associated with the transmit power constraint of the BS. The selection of  $\lambda_p$ can be set by hyperparameter optimization schemes, e.g., using the bisection method \cite{shi2011iteratively,pan2020multicell}. Algorithm~\ref{algorithm_digital} is ignorant of the fact that communications takes place in the near-field, as this property is only encapsulated in the equivalent channel vectors $\{\myVec{a}_m\}$. Nonetheless, as we show numerically  in Section~\ref{sec:Sims}, this optimization method, that targets the sum-rate and does not explicitly account for the resulting beam pattern, yields focused beams, which  allows multiple users to co-exist with minimal cross interference while residing in the same angular direction.

\vspace{-0.2cm}
\subsection{Beam Focusing via Phase-Shifters Based Hybrid Precoding}\label{sec:hybrid}
\vspace{-0.1cm} 
We next focus on the design of phase-shifters based hybrid antenna architecture.  In this case, by defining the $N_d\times M$ matrix $\myMat{W}\triangleq[\myVec{w}_1,\ldots,\myVec{w}_M]$, problem \eqref{eq:Original_optimization} is re-expressed as 
\begin{equation} \label{eq:optimization_problem_hybrid}
\begin{split}
&\max_{ \left\{{\bf w}_m\right\},{\bf Q}}~~\sum_{m=1}^{M} {\rm log}_{2}\left(1+\frac{\left|{\bf a}_m^H\, {\bf Q}{\bf w}_m \right|^{2}}{\sum_{j \neq m}\left|{\bf a}_m^H\, {\bf Q}{\bf w}_j \right|^{2}+\sigma^{2}}\right),\\
&~~~s.t.~~~~~~ \left\|\mathbf{Q} \mathbf{W} \right\|_{F}^{2} \leq P_{\rm max},~\mathbf{Q}\in \mathcal{F}^{N\times N_{\rm RF}}.
\end{split}
\end{equation} 
Problem \eqref{eq:optimization_problem_hybrid} is non-convex due to both the coupled optimization variables and the unit modulus constraints.  Intuitively, the  hybrid solution to problem \eqref{eq:optimization_problem_hybrid} should be sufficiently “close” to the  fully-digital solution of problem \eqref{eq:Original_optimization}. Therefore, our design seeks to identify the hybrid precoding mapping which is the closest to the resulting fully-digital precoding in the Frobenious norm sense. This strategy is quite often used for optimizing hybrid analog/digital systems, see, e.g., \cite{ioushua2019family,shlezinger2019dynamic,yu2016alternating, gong2019rf}. 
Specifically, let $\tilde{\bf W}_{\rm opt}=\left[\tilde{\bf w}_1,\cdots, \tilde{\bf w}_M\right]$ be the unconstrained precoding matrix obtained via Algorithm~\ref{algorithm_digital}. The resulting surrogate optimization problem is given by 
%
\ifsingle
\begin{equation} \label{eq:equivalent_problem}
\min_{ {\bf Q},{\bf W} }~~\left\|\tilde{\bf W}_{\rm opt}-{\bf Q}{\bf W}\right\|^2 
~~s.t.~~~\mathbf{Q} \in \mathcal{F}^{N\times N_{\rm RF}}.
\end{equation} 
\else 
\begin{equation} \label{eq:equivalent_problem}
\begin{split}
& \min_{ {\bf Q},{\bf W} }~~\left\|\tilde{\bf W}_{\rm opt}-{\bf Q}{\bf W}\right\|^2 \\
&~~s.t.~~~\mathbf{Q} \in \mathcal{F}^{N\times N_{\rm RF}}.
\end{split}
\end{equation} 
\fi
%

Note that we have temporarily neglected the power constraint in \eqref{eq:equivalent_problem}. Nonetheless, once $\myMat{Q}$ and $\myMat{W}$ are tuned to optimize \eqref{eq:equivalent_problem}, we  update the digital precoder by multiplying a factor, i.e., ${\bf W} =  \frac{\sqrt{P_{\rm max}}{\bf W}}{\left\|{\bf Q} {\bf W}\right\|^{2}}$. Thus, we tackle \eqref{eq:equivalent_problem} using alternating optimization. In particular, for a given $\myMat{Q}$, the digital precoding matrix $\myMat{W}$ which minimizes \eqref{eq:equivalent_problem} is stated in the following lemma:
\begin{lemma}
\label{lem:DigitalHBF}
For a given $\myMat{Q}$, the matrix $\myMat{W}$ which minimizes \eqref{eq:equivalent_problem} is
\begin{equation}\label{eq:digital_solution}
{\bf W} = \left({\bf Q}^H{\bf Q}\right)^{-1}{\bf Q}^H\tilde{\bf W}_{\rm opt}.
\end{equation}
\end{lemma}
\begin{IEEEproof}
The lemma is obtained as the least-squares solution to \eqref{eq:equivalent_problem} with fixed $\myMat{Q}$.
\end{IEEEproof}

To optimize $\myMat{Q}$ for a given $\myMat{W}$, we exploit the fact that the unit- modulus constraint on the norm of $\myMat{Q}$ bears similarity to constraints encountered in optimizing \acp{ris}. In particular, by defining  ${\bf q}_{\rm S}={\rm Vec}\left({\bf Q}\right)$, we equivalently reformulate \eqref{eq:equivalent_problem} for a given $\myMat{W}$ as
\begin{equation} \label{eq:analog_equivalent_vector}
\min_{{\bf q}_{\rm S} \in \mathcal{M}}~~f\left({\bf q}_{\rm S}\right) \triangleq  \left\|{\rm Vec}\left(\tilde{\bf W}_{\rm opt}\right)-\left({\bf W}^T \otimes {\bf I}\right){\bf q}_{\rm S}\right\|^2
\end{equation} 
where $\mathcal{S} = \left\{{\bf q}_{\rm S} \in {\mathbb{C}}^L:\left|{\bf q}_{{\rm S},1}\right|=\cdots=\left|{\bf q}_{{\rm S},L}\right|=1\right\}$ denotes the search space, and $L \triangleq N \times N_{\rm RF}$.

Note that the search space $\mathcal{S}$ in \eqref{eq:analog_equivalent_vector} is the product of $L$ complex circles, which is a Riemannian submanifold of ${\mathbb{C}}^L$. Therefore, following \cite{yu2019miso,guo2020weighted, pan2020multicell}, we tackle \eqref{eq:analog_equivalent_vector} using  the Riemannian conjugate gradient algorithm. The solution to \eqref{eq:analog_equivalent_vector} is thus updated based on the following formula
\begin{equation}\label{eq:step1}
  {\bf q}_{\rm S}^{t+1} = \mathcal{R}_t\left({\bf q}_{\rm S}^t + \varsigma^t { \boldsymbol \eta}^t\right),  
\end{equation}
where ${\bf q}_{\rm S}^t \in \mathcal{S}$ and ${\bf q}_{\rm S}^{t+1} \in \mathcal{S}$ denote the current point and the next point, respectively; $\varsigma^t > 0$ and 
${ \boldsymbol \eta}^t$ are the Armijo step size \cite{shewchuk1994introduction} and the search direction at the point ${\bf q}_{\rm S}^t$, respectively; $\mathcal{R}_t\left(\cdot\right)$ denotes the retraction operator, which projects the vector ${\bf q}_{\rm S}^t + \varsigma^t { \boldsymbol \eta}^t$ to the search space $\mathcal{S}$ via element-wise retraction, i.e., $ \left({\bf q}_{{\rm S}}^{t+1} \right)_{l} = \frac{\left({\bf q}_{\rm S}^t + \varsigma^t { \boldsymbol \eta}^t\right)_{l}}{\left|\left({\bf q}_{\rm S}^t + \varsigma^t { \boldsymbol \eta}^t\right)_{l}\right|}$ for $l =1,\cdots,L$. Let $T_{{\bf q}_{\rm S}^{t}} \mathcal{S}$ denote the tangent space of the complex circle manifold $\mathcal{S}$ at the point ${\bf q}_{\rm S}^t$, which is composed of all the tangent vectors that tangentially pass through ${\bf q}_{\rm S}^t$.
The search direction ${ \boldsymbol \eta}^t$ lies in $T_{{\bf q}_{\rm S}^{t}} \mathcal{S}$ \cite{absil2009optimization}
, given by
 \begin{equation}\label{eq:step2}
     { \boldsymbol \eta}^t = -  \operatorname{grad}\, f\left({\bf q}_{\rm S}^t\right) + \alpha^t \mathcal{T}_{{\bf q}_{\rm S}^{t-1} \rightarrow {\bf q}_{\rm S}^{t}}\left(\boldsymbol{\eta}^{t-1}\right),
 \end{equation}
 where  $\alpha^t$ is chosen as the Polak-Ribiere parameter \cite{shewchuk1994introduction}; $\operatorname{grad}\,f\left({\bf q}_{\rm S}^t\right)$ represents the Riemannian gradient of $f\left({\bf q}_{\rm S}\right)$ at the point ${\bf q}_{\rm S}^t$, which is obtained by orthogonally projecting 
 its Euclidean gradient, denoted by $\nabla f\left({\bf q}_{\rm S}^t\right)$,  onto the tangent space $T_{{\bf q}_{\rm S}^{t}} \mathcal{S}$, i.e.,
\begin{equation} \label{eq:projection_step1}
\begin{split}
\operatorname{grad}\,f\left({\bf q}_{\rm S}^t\right) &=\left[\begin{array}{c}
\left(\nabla f\left({\bf q}_{\rm S}^t\right)\right)_{1}-\operatorname{Re}\left\{\left(\nabla f\left({\bf q}_{\rm S}^t\right)\right)_{1} \times \left({\bf q}_{\rm S}^t\right)_{1}^{\dag}\right\} \left({\bf q}_{\rm S}^t\right)_{1} \\
\vdots \\
\left(\nabla f\left({\bf q}_{\rm S}^t\right)\right)_{L}-\operatorname{Re}\left\{\left(\nabla f\left({\bf q}_{\rm S}^t\right)\right)_{L} \times \left({\bf q}_{\rm S}^t\right)_{L}^{\dag}\right\} \left({\bf q}_{\rm S}^t\right)_{L}
\end{array}\right] \\
&=\nabla \,f\left({\bf q}_{\rm S}^t\right) - \operatorname{Re}\left\{\nabla\, f\left({\bf q}_{\rm S}^t\right) \circ ({{\bf q}_{\rm S}^t})^{\dag}\right\} \circ {{\bf q}_{\rm S}^t}\,,
\end{split}
\end{equation}
where $({\bf x})_{i}$ denotes the $i$th entry of a vector ${\bf x}$, $\operatorname{Re}\{\cdot\}$ denotes the real part of a complex number, and the notation $\circ $ represents the Hadamard production (element-wise multiplication) operation. The Euclidean gradient is defined as
$ \nabla\,f\left({\bf q}_{\rm S}^t\right) = 2 \left({\bf W}^{\dag} \otimes {\bf I}\right) \left(\left({\bf W}^T \otimes {\bf I}\right){\bf q}_{\rm S}^t - {\rm Vec}\left(\tilde{\bf W}_{\rm opt}\right) \right)$.

In \eqref{eq:step2},  $\mathcal{T}_{{\bf q}_{\rm S}^{t-1} \rightarrow {\bf q}_{\rm S}^{t}}\left(\boldsymbol{\eta}^{t-1}\right)$  denotes the vector transport, which maps the previous search direction $\boldsymbol{\eta}^{t-1}$ (lying in the tangent space $T_{{\bf q}_{\rm S}^{t-1}} \mathcal{S}$ ) to the tangent space $T_{{\bf q}_{\rm S}^{t}} \mathcal{S}$. As a result, $\operatorname{grad}\, f\left({\bf q}_{\rm S}^t\right)$ and $\mathcal{T}_{{\bf q}_{\rm S}^{t-1} \rightarrow {\bf q}_{\rm S}^{t}}\left(\boldsymbol{\eta}^{t-1}\right)$ lie in the same tangent space  $T_{{\bf q}_{\rm S}^{t}} \mathcal{S}$, and thus the sum operation in \eqref{eq:step2} makes sense.  The vector transport is given by
%
$\mathcal{T}_{{\bf q}_{\rm S}^{t-1} \rightarrow {\bf q}_{\rm S}^{t}}\left(\boldsymbol{\eta}^{t-1}\right) = \boldsymbol{\eta}^{t-1}-\operatorname{Re}\left\{\boldsymbol{\eta}^{t-1} \circ \left({\bf q}_{\rm S}^{t}\right)^{\dag}\right\} \circ {\bf q}_{\rm S}^{t}\,$.

Finally, the overall resulting configuration algorithm for hybrid analog precoders is summarized as Algorithm~\ref{algorithm_RGA}.
 \begin{algorithm}
 \caption{Alternating optimization of hybrid precoders.}
 \label{algorithm_RGA}
 \begin{algorithmic}[1]
  \renewcommand{\algorithmicrequire}{\textbf{Initialize:}} \REQUIRE  $\myMat{W}^0$, $\myMat{Q}^0$; \\
  \STATE Obtain $\tilde{\bf W}_{\rm opt}$ via Algorithm \ref{algorithm_digital};  \\
  \FOR{$j=1,\ldots,j_{\rm max}$} 
  \STATE ${\bf q}_{\rm S}^0={\rm Vec}\left({\bf Q}^0\right)$, ${ \boldsymbol \eta}^0 = -  \operatorname{grad}\, f\left({\bf q}_{\rm S}^0\right)$;\\
  \FOR{$t=1,\ldots,t_{\rm max}$}
     \STATE \textit{Find the next point ${\bf q}_{\rm S}^{t+1}$ according to \eqref{eq:step1}};
  \STATE \textit{Update the search direction ${ \boldsymbol \eta}^{t+1}$ according to \eqref{eq:step2}};
  \ENDFOR
  \STATE Set $\myMat{Q}^j = {\rm Vec}^{-1}\left(  {\bf q}_{\rm S}^{t_{\max}}\right)$; \\
  \STATE Set $\myMat{W}^j$ via Lemma~\ref{lem:DigitalHBF} with $\myMat{Q}=\myMat{Q}^j$.
  \ENDFOR
 \renewcommand{\algorithmicrequire}{\textbf{Output:}} \REQUIRE  $\myMat{Q}^j$,  $\myMat{W}^j$.
 \end{algorithmic} 
 \end{algorithm}
\vspace{-0.2cm}
\subsection{DMA-Based Beam Focusing}\label{sec:DMA}
\vspace{-0.1cm}
Here, we consider the configuration of DMAs for maximizing the sum-rate in near-field downlink communications. We note that the architectures considered in the previous subsections are relatively well-studied in the wireless communications literature, and thus we were able to utilize methods previously derived for similar setups to optimize the precoder. However, as the application of \acp{dma} for wireless communications is a relatively new area of research, 
in the following we derive a dedicated algorithm for configuring their weights based on \eqref{eq:Original_optimization}. 
In particular, we first reformulate \eqref{eq:Original_optimization} for the case of $\mathcal{W}=\mathcal{W}_{\rm DMA}$ as  
\begin{equation} \label{eq:optimization_problem_DMA}
\begin{split}
&\max_{ \left\{{\bf w}_m\right\},{\bf Q}}~~\sum_{m=1}^{M} {\rm log}_{2}\left(1+\frac{\left|{\bf a}_m^H\, \mathbf{H} \mathbf{Q}\, {\bf w}_m \right|^{2}}{\sum_{j \neq m}\left|{\bf a}_m^H\, \mathbf{H} \mathbf{Q}\, {\bf w}_j \right|^{2}+\sigma^{2}}\right)\\
&~~s.t.~~~~~~\eqref{eq: weighting_matrix}, \quad q_{i, l} \in \mathcal{Q}, \forall i,l, \quad \sum_{m=1}^{M} \left\|{\bf w}_m\right\|^2 \leq P_{\rm max}.
\end{split}
\end{equation} 	
We note that \eqref{eq:optimization_problem_DMA} is slightly different from \eqref{eq:Original_optimization}, as the power constraint here is imposed on the digital output, and not the transmitted signal. Nonetheless, as discussed in the previous subsection, one can derive the overall system based on \eqref{eq:optimization_problem_DMA}, and scale the digital precoder such that the power constraint in \eqref{eq:Original_optimization} is satisfied. 
As the joint design of the configuration of the DMA weights along with the digital precoding vector based on the non-convex problem \eqref{eq:optimization_problem_DMA} is challenging, we begin by considering a single-user setup to gain more design insights. 
After that, we extend our study to the multi-user case of $M>1$, and propose an alternating algorithm to deal with the resulting non-convex optimization problem.

\subsubsection{Single-user case} \label{sec: single-user}

For the single-user case, there 
is no inter-user interference. Consequently, the achievable rate   is given by
$R =  {\rm log}_{2}\left(1+\frac{1}{ \sigma^{2}}\left|{\bf a}^H\, \mathbf{H} \mathbf{Q}\, {\bf w} \right|^{2}\right)$,
%
where we have dropped the user index subscript $m$. 
 Due to the monotonicity of the logarithm function, the  optimization problem \eqref{eq:optimization_problem_DMA} is  equivalently rewritten as 
 \ifsingle
 \begin{equation} \label{eq:optimization_problem_single}
 \max_{ {\bf w},{\bf Q}}~~\left|{\bf a}^H\, \mathbf{H} \mathbf{Q}\, {\bf w}\right|^{2}, ~~s.t.~~~~\eqref{eq: weighting_matrix},\quad q_{i, l} \in \mathcal{Q}, \forall i,l, \quad \left\|{\bf w}\right\|^2 \leq P_{\rm max}.
 \end{equation} 
 \else
 \begin{equation} \label{eq:optimization_problem_single}
 \begin{split}
 &\max_{ {\bf w},{\bf Q}}~~\left|{\bf a}^H\, \mathbf{H} \mathbf{Q}\, {\bf w}\right|^{2}\\
 &~~s.t.~~~~\eqref{eq: weighting_matrix},\quad q_{i, l} \in \mathcal{Q}, \forall i,l, \quad \left\|{\bf w}\right\|^2 \leq P_{\rm max}.
 \end{split}
 \end{equation} 
 \fi

\begin{figure}
		\centering	
	\includegraphics[width=0.5\columnwidth]{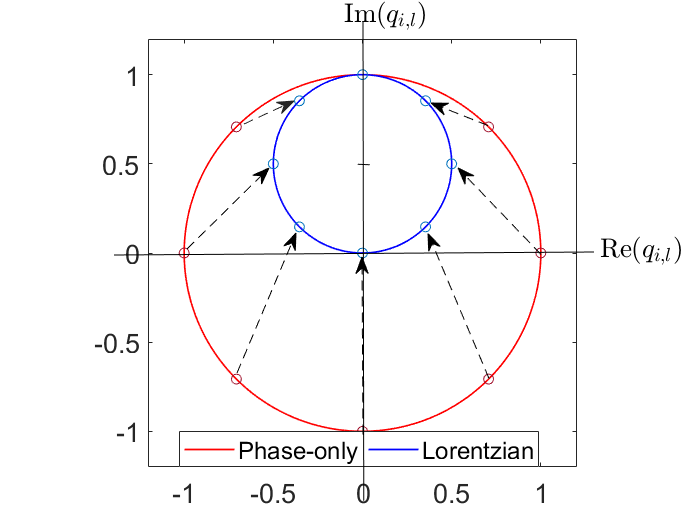}
		\vspace{-0.4cm}
		\caption{Illustration of the phase-only weights (outer circle) and Lorentzian weights (inner circle) in the complex plane.  Arrows indicate the mapping between
the phase-only weights and Lorentzian weights points.} 
		\label{fig:mapping}
	\end{figure}

Although \eqref{eq:optimization_problem_single} is notably simpler to tackle compared to \eqref{eq:optimization_problem_DMA}, it still involves coupled optimization variables in the objective function, as well as  the non-trivial element-wise constraint $\mySet{Q}$. Specifically, each element response  $q_{i,l}$ should take the Lorentzian-constrained form in \eqref{eqn:FreqSel}, which is represented by a circle in the top half of the complex plane (inner circle in Fig. \ref{fig:mapping}). Thus, the phase and amplitude of $q_{i,l}$ are coupled, which makes it  challenging to solve \eqref{eq:optimization_problem_single}. To tackle this problem, we employ the constant amplitude approach proposed in \cite{smith2017analysis} to relax the Lorentzian constraint to the phase-only weights constraint with constant amplitude and arbitrary phase,  given by 
$q_{i,l} \in \mySet{F}$ as defined in \eqref{eqn:analog_constraint}, for each $i,l$.
%
%
The feasible set $ \mySet{F}$ is a circle with unit radius and centered at the origin (outer circle in Fig. \ref{fig:mapping}). 
%

We next focus on  problem \eqref{eq:optimization_problem_single} with  $\mathcal{Q}$ replaced by $\mySet{F}$, and then use the solution to tune the Lorentzian-constrained weights via projection as illustrated in Fig. \ref{fig:mapping}. Due to the coupled optimization variables, the problem is still non-convex. Nonetheless, we are able to solve it in closed-form, as stated in the following theorem.

\begin{theorem}
\label{thm:SingleUser}
Let $({\bf Q}^*,{\bf w}^*)$ be the solution to  \eqref{eq:optimization_problem_single}  with  $\mathcal{Q} =  \mySet{F}$. According to the structure constraint \eqref{eq: weighting_matrix}, each non-zero element of ${\bf Q}^*$ is $q_{i, l}^* = e^{ \jmath \psi_{i,l}^*}$, with $\psi_{i,l}^* =k|\p_m-\p_{i,l}| + \beta_i \rho_{i,l}$, and ${\bf w}^*=\frac{\sqrt{P_{\rm max}}\left({\bf a}^H\, \mathbf{H} \mathbf{Q}^*\right)^H}{\left\|{\bf a}^H\, \mathbf{H} \mathbf{Q}^*\right\|}$.
\end{theorem}
\ifFullVersion
\begin{IEEEproof}
The proof is given in Appendix \ref{app:Proof1}.  
\end{IEEEproof}
\fi
 
From Theorem~\ref{thm:SingleUser}, we can see that  the optimized  phase $\psi_{i,l}^*$ of each metamaterial element includes two parts: The first is  the term $k|\p-\p_{i,l}|$ for achieving near-field focusing,  i.e., coherent sum of signal components in location $\p$; the other part  compensates for the transmission delay in the microstrip, given by $\beta_i \rho_{i,l}$ \eqref{eq:delay_in_microstrip}, such that the signals are transmitted synchronously. It is noted that a similar phase profile would be found using an holographic design process \cite{gowda2018focusing}.

As the  weighting  $ q_{i,l}^*=e^{ \jmath \psi_{i,l}^*}$ does not satisfy the Lorentzian form defined in \eqref{eqn:FreqSel}, we project it onto $\mySet{Q}$ as illustrated in Fig.~\ref{fig:mapping}. In particular, the resulting non-zero weights are given by ${\hat q}_{i,l}=\frac{j+e^{j \psi_{i,l}^*}}{2}$ \cite{smith2017analysis}, and thus ${\hat q}_{i,l}\in \mathcal{Q}$. Hence, the resulting $\hat{\myMat{Q}}$  can be used as an approximate solution to the original optimization problem \eqref{eq:optimization_problem_single}. The numerical results provided in  Section~\ref{sec:Sims} verify that deriving the elements response assuming weights of the form $\mySet{F}$ followed by their projection onto $\mathcal{Q}$ yields accurate focused beams. 
The proposed relaxation and projection approach facilitates the design of effective DMA weights for the single-user case. 

\subsubsection{Multi-user case} \label{sec:multi-user}

Next, we study the formulation of focused beams using DMAs for multi-user setups by tackling   \eqref{eq:optimization_problem_DMA} with $M > 1$.
Following the strategy used in Section~\ref{sec:hybrid}, we   optimize $\myMat{Q}$ and $\{\myVec{w}_m\}$ in an alternating manner, while relaxing the constraints on the feasible  elements response as done in the single-user setup. 
 %
%
 %
 %
%
This procedure is iterated until convergence. In the following, we show how to solve \eqref{eq:optimization_problem_DMA} for fixed $\myMat{Q}$ and for fixed $\{\myVec{w}_m\}$, respectively.
 
(a) \textit{{Solving \eqref{eq:optimization_problem_DMA} w.r.t. $\{\myVec{w}_m\}$}:} When $\bf Q$ is fixed, the form of problem \eqref{eq:optimization_problem_DMA} is similar to the  sum-rate maximization problem  studied in the previous Section \ref{sec:fully_digital}. Hence, following Lemma~\ref{Lemma1}, 
problem \eqref{eq:optimization_problem_DMA} with fixed $\myMat{Q}$ is equivalent to  
\begin{equation} \label{eq:equivalent_subproblem_w}
\begin{split}
&\max_{ \left\{{\bf w}_m, u_m, v_m\right\}}~~~\sum_{m=1}^{M}~{\rm log}_2 (v_m) - v_m e_m^{\rm DMA} \left(u_m,\left\{{\bf w}_m\right\}\right)\\
&~~~~~~~s.t.~~~~~~~~\sum_{m=1}^{M} \left\|{\bf w}_m\right\|^2 \leq P_{\rm max}, \quad v_m \geq 0,~  m\in \mySet{M}.
\end{split}
\end{equation} 
where $u_m$ and $v_m$ are auxiliary variables, and 
$e_m^{\rm DMA}\left(u_m,\left\{{\bf w}_m\right\}\right) = \left|1-u_m\, {\bf a}_m^H\, \mathbf{H} \mathbf{Q}\, {\bf w}_m \right|^2 \notag +   \sum_{j \neq m}\left|u_m\,{\bf a}_m^H\, \mathbf{H} \mathbf{Q}\, {\bf w}_j\right|^{2}  + \sigma^2\,\left|u_m\right|^2$. 
One can verify that problem \eqref{eq:equivalent_subproblem_w} has the same structure as problem \eqref{eq:problem_digital_w}. Hence, Algorithm \ref{algorithm_digital} is applied to solve \eqref{eq:equivalent_subproblem_w}.

\textit{(b) {Solving \eqref{eq:optimization_problem_DMA} w.r.t. $\myMat{Q}$}: }
%
 %
 %
To proceed, we first define the $ N_d^2 \cdot N_e  \times 1$ vectors  ${\bf q}={\rm Vec}\left(\bf Q \right) $, and ${\bf z}_{j,m}=\left({\bf w}_j^T \otimes ({\bf a}_m^H\, \mathbf{H})\right)^H$. Using these definitions, we then identify an equivalent optimization problem, as stated in following theorem.
\begin{theorem}
\label{thm:multi-user}
Problem \eqref{eq:optimization_problem_DMA} with fixed $\{{\bf w}_m\}$ is equivalent to the following problem:
\begin{equation} \label{eq:simplifiedx}
\begin{split}
&\max_{ {\bf \bar q}}~~f\left({\bf \bar q}\right) \triangleq  \sum_{m=1}^{M} {\rm log}_{2}\left(1+\frac{\left|{\bf \bar q}^H{\bf \bar z}_{m,m} \right|^{2}}{\sum_{j \neq m}\left|{\bf \bar q}^H{\bf \bar z}_{j,m}\right|^{2}+\sigma^{2}}\right)\\
&~~s.t.~~~{\bar q_{l}} \in \mathcal{Q},~ l \in  \mathcal{A}_q,
\end{split}
\end{equation} 
where $\mathcal{A}_q$ is the set of all non-zero elements of ${\bf q}$, ${\bf \bar q}$ is the modified version of ${\bf q}$ obtained by removing all the zero elements of  ${\bf q}$; ${\bf \bar z}_{j,m}$ is the modified version of ${\bf z}_{j,m}$, which is obtained by removing the elements having the same index as the zero elements of ${\bf q}$.
\end{theorem}
\ifFullVersion
\begin{IEEEproof}
The proof is given in Appendix \ref{app:Proof2}.  
\end{IEEEproof}
\fi

\smallskip
The equivalence
between the relaxed \eqref{eq:optimization_problem_DMA} and  \eqref{eq:simplifiedx} holds in the sense that they have the same optimal value, and the  solution to the relaxation of \eqref{eq:optimization_problem_DMA} can be recovered from the  solution to  \eqref{eq:simplifiedx} according to \eqref{eq: weighting_matrix}. 
Although  \eqref{eq:simplifiedx} is still non-convex, we can find its approximate solution using alternating optimization, by iteratively optimizing each element of ${\bf \bar q}$ separately while keeping the remaining elements fixed. The resulting optimization problem for the $l$th element is 
\begin{equation} \label{eq:single_variable} 
\max_{ 0 \leq \phi_l \leq 2\pi}~~f\left({\bf \bar q}\left( \phi_l\right)\right)\triangleq \sum_{m=1}^{M} {\rm log}_{2}\left(1+\frac{\left|{\bf \bar q}\left( \phi_l\right)^H{\bf \bar z}_{m,m} \right|^{2}}{\sum_{j \neq m}\left|{\bf \bar q}\left( \phi_l\right)^H{\bf \bar z}_{j,m}\right|^{2}+\sigma^{2}}\right), 
\end{equation} 
where 
${\bf \bar q}\left( \phi_l\right) \triangleq \Big[{\bar q}_1,\cdots,{\bar q}_{l-1}, \frac{j+e^{j \phi_l}}{2}, {\bar q}_{l+1},\cdots,{\bar q}_{N_d N_e}\Big], l \in  \mathcal{A}_q$.

Problem \eqref{eq:single_variable} is a single variable optimization problem  with respect to $\phi_l$. Although it is challenging to find a closed-form solution for \eqref{eq:single_variable}, we can easily solve it by using numerical techniques such as a one-dimensional search.
The proposed algorithm for solving  \eqref{eq:optimization_problem_DMA} for the case of $M>1$ is summarized as Algorithm \ref{algorithm3}.
In  Algorithm \ref{algorithm3}, steps \ref{stp:Loop1Start} to \ref{stp:Loop1End} solve  \eqref{eq:equivalent_subproblem_w} for a fixed ${\bf Q}$ using the similar procedure as in Algorithm \ref{algorithm_digital}, and steps \ref{stp:Loop2Start} to \ref{stp:Loop2End} solve \eqref{eq:simplifiedx} for a fixed $\left\{{\bf w}_m\right\}$ using the alternating optimization approach.

 \begin{algorithm}
 \caption{Alternating optimization of DMA precoders for $M >1$.}
 \setstretch{1}
 \label{algorithm3}
 \begin{algorithmic}[1]
 \renewcommand{\algorithmicrequire}{\textbf{Initialize:}}
 \REQUIRE $\left\{{\bf w}_m^0\right\}_{m=1}^M$, ${\bf Q}^0$;
 \FOR{$t = 1,\ldots,t_{\max}$}
 \STATE {Update ${\bf g}_m \triangleq ({\bf Q}^{t-1})^H{\bf H}^H{\bf a}_m, \forall m$}
 \FOR{$t_1 = 1,\ldots,t_{\max}$ \label{stp:Loop1Start}}
  \STATE Update $u_m^{t_1} = \frac{{\bf g}_m^H\, {\bf w}_m^{{t_1}-1}}{\sum_{j=1}^M \left|{\bf g}_m^H\,{\bf w}_j^{{t_1}-1}\right|^{2} + \sigma^2},~ \forall m$;
  \STATE Update $v_m^{t_1} = (e_m^{\rm DMA}\left(u_m^{t_1},\left\{{\bf w}_m^{{t_1}-1}\right\}\right))^{-1},~\forall m$;
  \STATE Update ${\bf w}_m^{t_1} \!=\! u_m^{t_1}\,v_m^{t_1}\,\left( \sum_{j=1}^M\,v_j^{t_1}\,\left|u_j^{t_1}\right|^2{\bf g}_j{\bf g}_j^H \!+\! \lambda_p {\bf I} \right)^{-1}{\bf g}_m, \forall m$;
 \ENDFOR  \label{stp:Loop1End}
 \STATE Update ${\bf w}_m={\bf w}_m^{t_1}, \forall m$;
 \STATE Update ${\bf \bar z}_{j,m}, \forall m,j$;
 \STATE Update ${\bf \bar q}={\rm Vec}\left({\bf Q}^{t-1} \right)$;
 \FOR{$l = 1,\ldots,N_d N_e$ \label{stp:Loop2Start}}
\STATE {Update} $\phi_l$ by solving problem \eqref{eq:single_variable};
  \STATE {Update the $l$th element of ${\bf \bar q}$}: ${ \bar q}_l=\frac{j+e^{j \phi_l}}{2}$;\\
 \ENDFOR  \label{stp:Loop2End}
\STATE Update ${\bf Q}^{t}$ with non-zero entries taken from ${\bf \bar q}$;
\ENDFOR
\renewcommand{\algorithmicensure}{\textbf{Output:}}
\ENSURE   $\{\myVec{w}_m^{t_1}\}$ and ${\bf Q}^{t}$.
\end{algorithmic}
\end{algorithm}


\vspace{-0.2cm}
\subsection{Discussion}
\label{sec:Discussion}
\vspace{-0.1cm}
In line with traditional multi-user communication schemes, our derivation in the previous sections considers the achievable sum-rate as the objective function to be optimized. Consequently, while our work focuses on beam focusing, we do not explicitly design the transmission beam patterns to generate focused beams, but rather to establish reliable high-rate communications. The fact that we are operating in the near-field, from which the beam focusing ability arises, is implicitly encapsulated in the objective via the vectors $\{{\bf a}_m\}$, as discussed in Section~\ref{sec:model}. This property allowed us to combine in our derivations methods developed for far-field systems, as discussed in, e.g., Section~\ref{sec:fully_digital}. While we do not directly enforce the generation of focused beams, our numerical study in Section \ref{sec:Sims} demonstrates that such beams are indeed generated when seeking to optimize the sum-rate, enabling users with identical angles to simultaneously achieve high rates with little interference.  

Among the considered architectures, the fully-digital antenna supports the most flexible design, and it is expected to achieve the largest sum-rates among all considered architectures for a given \ac{upa} with fixed element placing. However, such architectures assign a costly RF chain to each element, and thus may be prohibitively costly when the number of antennas elements $N$ becomes very large. To achieve cost-effective power transmission, hybrid antenna architectures with limited RF chains are often adopted.  Nonetheless, as the total number of required phase shifters is very large, the power consumption of active phase-shifters based analog precoding may also become significant. 
The third considered architecture, i.e., DMAs, is the most scalable in terms of cost and power efficiency. Nonetheless, it is also the most challenging to design due to the Lorentzian form of its elements, whose gain and phase are coupled. For this reason, our design method resorted to optimizing each element separately in an alternating manner, as detailed in Algorithm~\ref{algorithm3}. 
Furthermore, DMAs are typically utilized with sub-wavelength element spacing, allowing to pack a larger number of elements in a given physical area compared to conventional antennas based on, e.g., patch arrays \cite{Akyildiz-2016NCN}. Therefore, for a given antenna aperture, DMAs are in fact capable of achieving the most focused beams among all considered architecture for the single-user case, as will be shown by the numerical results in Section~\ref{sec:Sims}.

The model used for the frequency response of the \ac{dma} elements is the Lorentzian-phase constrained form \eqref{eqn:FreqSel}, which does not vary in frequency.   Nonetheless, \ac{dma} elements can also be configured to exhibit controllable frequency selectivity, i.e., a different response can be configured in each frequency bin in a coupled manner \cite{wang2019dynamic2}. This advanced frequency selective analog signal processing capabilities, which are not available in conventional hybrid antenna architectures based on phase shifters \cite{ioushua2019family}, give rise to the possibility of generating frequency-variant beam focusing patterns,  facilitating high-rate wideband communications with a large number of users. Furthermore,  we focus on the typical \ac{dma} architecture comprised of a set of one-dimensional microstrips, which result in the equivalent partially-connected model in \eqref{eq: weighting_matrix} and hence brings in some performance loss compared to fully-connected analog combiners for the multi-user scenario.
However, one can also use two-dimensional waveguides \cite{imani2018two} resulting in an equivalent fully connected model, i.e., $\myMat{Q}$ is not restricted to take the form  \eqref{eq: weighting_matrix}, although such an architecture results in a more complex model for the propagation inside the waveguuide.  We leave these extensions for future work.

	\vspace{-0.2cm}
	\section{Numerical Evaluations}
	\label{sec:Sims}
	\vspace{-0.1cm}

Here, we provide numerical results to verify the near-field beam focusing capability of our proposed designs under three different antenna architectures.  We first consider the single-user scenario in Section \ref{subsec:single-user}, which demonstrates the gains of near-field beam focusing over beam steering in enhancing the signal strength of the target point. Then, in Section \ref{subsec:multi-user}, we show the advantage of beam focusing to distinguish different users for  multi-user communications.

Throughout the experimental study, we consider a planar array positioned in the $xy$-plane, with  carrier frequency set to $f_c=28$~GHz ($\lambda = 1.07$~cm). To make a fair comparison, We consider that the three different types of antennas have the same aperture and are all equal to $D=\sqrt2L$, with $L$ denoting antenna length. We consider $\lambda/2$ antenna separation for fully-digital and hybrid antennas, and $\lambda/5$ spacing between DMA elements within the same row unless other wise stated. The separation between rows could still be $\lambda/2$. Hence, the number of rows and elements in each row for the fully-digital and hybrid antennas are $N_d = N_e = \lfloor 2L/\lambda\rfloor$, where $\lfloor \cdot\rfloor$ is the integer floor function. For the phase-shifter based analog precoder we set the number of RF chains to $N_{\rm RF} = N_d$, while for the DMA, the number of microstrips and that of metamaterial elements are $N_d = \lfloor 2L/\lambda\rfloor$ and $N_e = \lfloor 5L/\lambda\rfloor$ respectively. We use   { $\alpha = 0.6~[{\rm m}^{-1}]$ and $\beta = 827.67  ~[{\rm m}^{-1}]$} to represent the propagation inside the DMA waveguides, assuming a microstip implemented in Duroid 5880 with 30 mill thickness. In addition, 
we set the maximum transmit power to $P_{\rm max} = -13\,$dBm,  and the noise power to $\sigma^2=-114\,$dBm\footnote{This setting corresponds to the noise power spectrum density at users is $-174$~dBm/Hz and signal bandwidth is $120\,$KHz, assuming the noise figure of each user to be $9$~dB. 

}.

\vspace{-0.2cm}
\subsection{Single-User Scenario}
\label{subsec:single-user}
\vspace{-0.1cm} 

\ifsingle
\begin{figure}
	\centering	
	\begin{minipage}{0.45\textwidth}
		\centering
		\scalebox{0.48}{\includegraphics{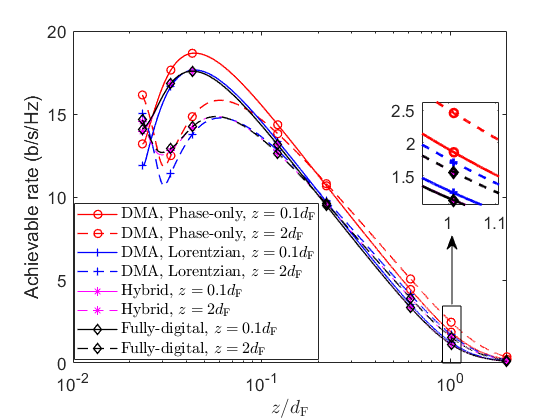}}
		\vspace{-0.4cm}
		\caption{Achievable rates of beam focusing at ${\rm F}_{\rm near}$ and beam steering at ${\rm F}_{\rm far}$.}
		\label{fig:rate_user1}		
	\end{minipage}
	$\quad$
	\begin{minipage}{0.45\textwidth}
		\centering
		\scalebox{0.48}{\includegraphics{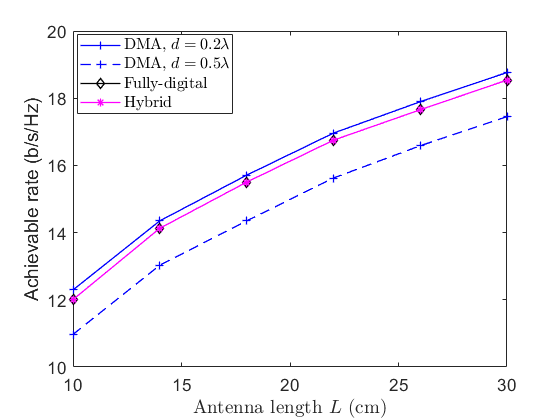}}
		\vspace{-0.4cm}
		\caption{Achievable rate versus the antenna length for focusing point at ${\rm F}_{\rm near}((x,y,z)=(0,0,150\, \lambda))$.}
		\label{fig:antenna_size}
	\end{minipage} 
\end{figure}
\else
	\begin{figure}
		\centering	
		\includegraphics[width=0.5\columnwidth]{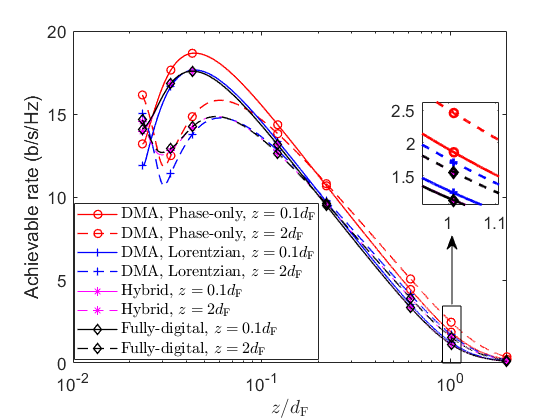}
		\vspace{-0.4cm}
		\caption{Achievable rates of beam focusing at $F_{\rm near}$ and beam steering at $F_{\rm far}$.}
		\label{fig:rate_user1}
	\end{figure}

	\begin{figure}
		\centering	
		\includegraphics[width=0.50\columnwidth]{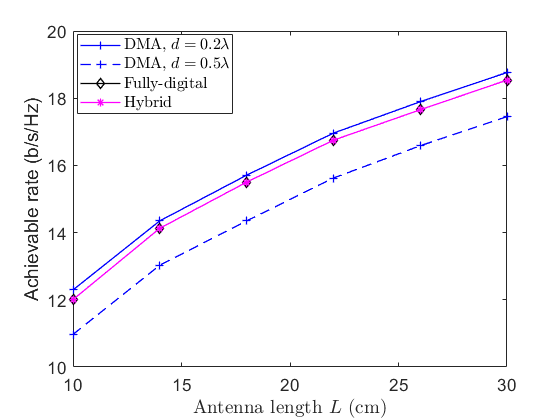}
		\vspace{-0.4cm}
		\caption{Achievable rate versus the antenna length. }
		\label{fig:antenna_size}
	\end{figure}
\fi
	
We first show the beam focusing design in the single-user scenario, where the antenna length is set to be $L=30$~cm. Fig. \ref{fig:rate_user1} illustrates the achievable rates along the $z$-axis direction, namely, when the single user is located in each specific point in the $z$-axis, achieved by the beam focusing and beam  steering solutions under the three different considered antenna architectures. For DMAs, we depict here the rate achieved with phase-only elements in addition to that achieved using their true Lorentizan, as phase-only DMA weights are derived as an intermediate step in the optimization algorithm detailed in Section~\ref{sec:DMA}. Beam focusing is obtained by setting the focusing points at the near-field ${\rm F}_{\rm near}((x,y,z)=(0,0,0.1\dF))$, whereas the beam  steering is at the far-field ${\rm F}_{\rm far}((x,y,z)=(0,0,2\dF))$. In Fig. \ref{fig:rate_user1}, we clearly observe that for all antenna architectures, near-field focusing can increase the signal strength, which in turn improves the achievable rate, when the user is located in the proximity of the focusing point of $z= 0.1 \dF$. Secondly, when the user is located in the direction of the focusing point but at a different distance from the antenna plane, the observed radiation is notably reduced compared with the corresponding far-field beam steering design, resulting in lower rates. 

The achievable rate of beam focusing in Fig.~\ref{fig:rate_user1} is much larger than that of the beam steering solution at the near-field focusing point ${\rm F}_{\rm near}$. In particular, the DMA architecture achieves higher rates compared to fully-digital and hybrid antennas. This is because DMA stacks more antenna elements within the same antenna area due to the smaller element spacing than the other two types of antennas. Moreover, we note that the hybrid antenna achieves the same performance as that of the fully-digital antenna. This observation is consistent with existing results in far-field communications, where it is established that any fully-digital beamforming solution can be realized by a hybrid beamforing solution when the number of RF chains is at least twice the number of users (data streams) \cite{sohrabi2016hybrid}. Finally, we note that the achievable rate of the  Lorentzian weights constraint is comparable to that of the phase-only weights constraints, which verifies the effectiveness of our DMA configuration approach. In the remainder of the experimental study we thus consider only DMAs with lorentzian-constrained weights. 

Next, Fig. \ref{fig:antenna_size} illustrates the effect of antenna size (or antenna aperture) on the achievable rate of each of the three antenna architectures. We set the focusing point at ${\rm F}_{\rm near}((x,y,z)=(0,0,150\, \lambda))$.
From Fig. \ref{fig:antenna_size}, we observe that as the antenna length $L$ increases, the achievable rate substantially improves for each of the three antenna architectures. This reveals an important insight that large antenna arrays can not only increase the near-field region ($d_\mathrm{F}=\frac{2\,D^2}{\lambda}$), but also enhance the signal strength at the near-field focusing point. Moreover, it is observed that the achievable rate of DMAs with $0.2\, \lambda$ element spacing is higher than that of both fully-digital and hybrid antennas, while the performance of DMAs with  $0.5\, \lambda$ element spacing is worse than the other two antenna architectures. This is expected since the smaller the element spacing, the stronger the antenna superdirectivity effect caused by antenna densification.


\vspace{-0.2cm}
\subsection{Multi-User Scenario}
\label{subsec:multi-user}
\vspace{-0.1cm}

\begin{figure} 
  \centering 
  \subfigure[Near-field: $z_1 = 0.1 \dF$  and $z_2 = 0.4 \dF$.]{ 
    \label{fig:subfig:near-field}
    \includegraphics[width=2.8in]{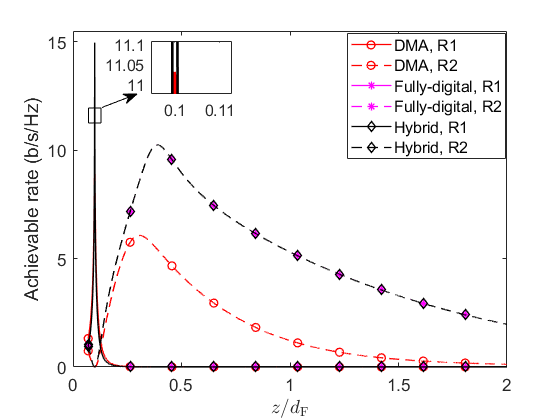} 
  } 
  \subfigure[Far-field: $z_1 = 1.5\dF$  and $z_2 = 1.8 \dF$]{ 
    \label{fig:subfig:far-field} 
    \includegraphics[width=2.8in]{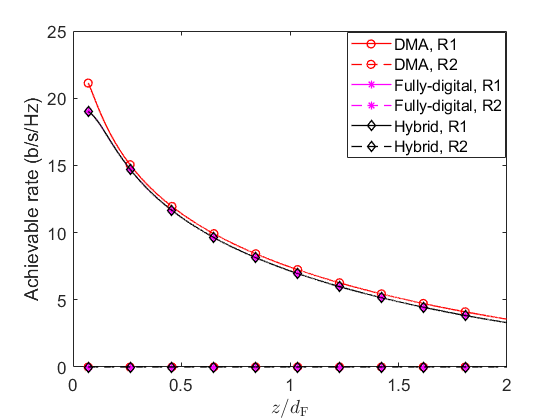}
    } 
  \caption{Achievable rates per user versus location along the $z$-axis.} 
  \label{fig:rate_twoUser} 
\end{figure}

	\begin{figure} 
  \centering 
    \subfigure[The normalized signal power of user 1]{ 
    \label{fig:subfig:user1 of Two-user case}
    \includegraphics[width=2.8in]{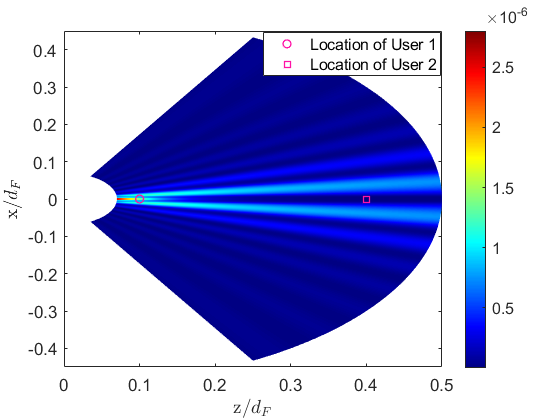} 
  } 
  \subfigure[The normalized signal power of user 2]{ 
    \label{fig:subfig:user2 of Two-user case}
    \includegraphics[width=2.8in]{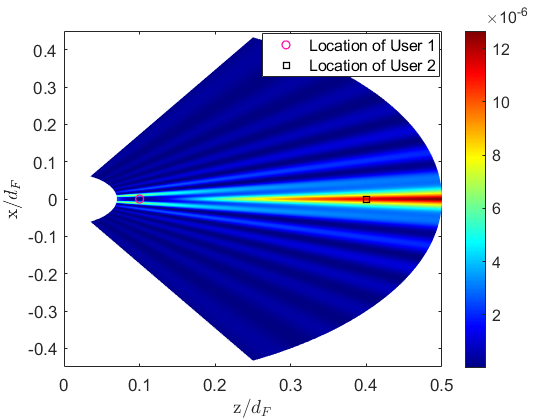} 
  } 
   \caption{The normalized signal power measurement of two users.}    \label{fig:far-field two_user} 
\end{figure}

\begin{figure}
		\centering	
		\includegraphics[width=0.50\columnwidth]{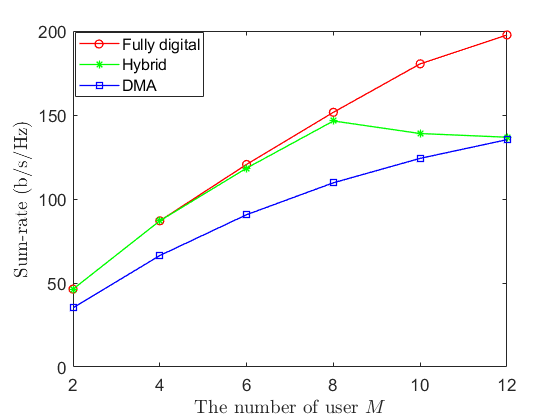}
		\vspace{-0.4cm}
		\caption{Achievable sum-rate versus the number of users. }
		\label{fig:sum-rate-multiuser}
	\end{figure}

We proceed to  show the advantage of beam focusing to distinguish different users for downlink communications. Having demonstrated the effect of antenna length on the achievable rate in the previous section, we fix the antenna length here to $L = 10~{\rm cm}$. We first study the two-user case in Figs. \ref{fig:rate_twoUser} and \ref{fig:far-field two_user}. Then we study the general multiple users ($M>2$) case where the users are randomly distributed on the $xz$-plane within the near-field region in Fig.~\ref{fig:sum-rate-multiuser}.

Fig. \ref{fig:rate_twoUser} illustrates the achievable rates of each of the two users, i.e. $\{R_1, R_2\}$, measured along  the $z$-axis. The two focal points of the users are located at: $(a)$ near-field region:  ${\rm F}_1((x,y,z_1)=(0,0,0.1\dF))$ and ${\rm F}_2((x,y,z_2)=(0,0,0.4  \dF))$; and (b) far-field region: ${\rm F}_1((x,y,z_1)=(0,0,1.5 \dF))$ and ${\rm F}_2((x,y,z_2)=(0,0,1.8 \dF))$.  
From Fig. \ref{fig:subfig:near-field}, it is observed that for all the studied  antenna architectures, the peak achievable rates of each of the two users occur when they are located around their corresponding focal points, implying that the designed focused beams for each type of the antennas are all capable of yielding reliable communications with minimal degradation due to interference. 
For example, the achievable rate of user 1 is maximized when it is located at $(0,0,z_1)$.  
Moreover, it is observed that the fully-digital and hybrid antennas achieve higher rates than the DMA architecture for the near-filed two-user scenario. This is due to the fact that we consider the DMA with one-dimensional microstrips, which results in a partially-connected analog combiner and hence brings in some performance loss compared to fully-connected analog combiners under the multiuser scenario. When utilizing conventional beam steering based on far-field assumptions, it is observed in Fig. \ref{fig:subfig:far-field} that user 2 achieves negligible rates, i.e., $R_2 \approx 0$, regardless of its distance from the transmit antenna. This is because, for far-field communications, conventional beam steering is unable to distinguish the two users with the same angular direction. Hence, in order to maximize the sum-rate,  conventional beam steering  allocates essentially all the transmit power to one user that has the better channel (i.e., smaller path loss). 

 In order to more explicitly illustrate the distinguishing ability of near-field beam focusing, in Fig. \ref{fig:far-field two_user} we further show the normalized signal power of the signal transmitted to each user along the whole $xz$-plane, using a fully-digital antenna. The received signal power of each user at each point is normalized by the corresponding channel gain to remove the effect of the distance between the user and the BS. Fig.~\ref{fig:subfig:user1 of Two-user case} shows the normalized signal power of user 1 at each point of near-field $xz$-plane, from which we can clearly see that the maximum normalized signal power is achieved at around the focusing point of user 1, while the minimum normalized signal power is achieved at around the focusing point of user 2. This result verifies our conclusion that near-field focusing can not only enhance the signal strength at the focusing point, but also eliminate the co-channel interference to other users, even if the two users lies in the same angular direction. The same is observed in Fig.~\ref{fig:subfig:user2 of Two-user case}, which illustrates the normalized signal power of user 2 along the near-field $xz$-plane.  From Figs. \ref{fig:rate_twoUser} and \ref{fig:far-field two_user}, we conclude that our proposed near-field beam focusing design allows to simultaneously communicate with multiple users located at the same angular direction, while such separation ability is not achievable for conventional beam steering. 

Finally, we study a more general scenario with different number of users randomly located at the near-field $xz$-plane. Fig. \ref{fig:sum-rate-multiuser} shows the achievable sum-rate versus the number of users $M$.
Particularly, we successively add users to the near-field $xz$-plane. 
From Fig. \ref{fig:sum-rate-multiuser}, it is observed that the achievable sum-rate of the three different antenna architectures all monotonically increase with the number of users $M$ when $M$ is small. However, as the number of users $M$ increases, the growth trend of the sum-rate achieved by both fully-digital and DMA architectures becomes slower. This is because as the number of users $M$ increases, the co-channel interference among users limits the increase of achievable sum-rate. On the other hand, it is observed that the achievable sum-rate of hybrid antenna starts to decrease when the number of users $M \geq 8$. This is because as the number of RF chains is constant, the hybrid beamforming solution cannot perfectly realize the optimal digital beamforming solution when the number of users is greater than half the number of RF chains.

	\vspace{-0.2cm}
	\section{Conclusions}
	\label{sec:Conclusions}
	\vspace{-0.1cm}
In this work, we studied the potential of beam focusing for a near-field multi-user MIMO communication scenario. We first provided the mathematical model for the near-field wireless channels and the transmission pattern for the three different types of antenna architectures, including fully-digital,  (phase shifters based-)  hybrid and  DMA architectures. We then formulated a near-field beam focusing problem for maximizing the achievable sum-rate. After that, we proposed efficient solutions based on the sum-rate maximization task for each of the antenna architectures.  Numerical results demonstrated that beam focusing results in an improved achievable rate in the near-field, and also has the potential of decreasing co-channel interference in multi-user communication scenarios. 
In particular, it is shown that the achievable sum-rate of hybrid architectures and DMAs is comparable to that of fully-digital architectures.

\ifFullVersion
\vspace{-0.2cm}
\begin{appendix}
	\numberwithin{proposition}{subsection} 
	\numberwithin{lemma}{subsection} 
	\numberwithin{corollary}{subsection} 
	\numberwithin{remark}{subsection} 
	\numberwithin{equation}{subsection}	
	%
	\vspace{-0.2cm}
	\subsection{Proof of Theorem \ref{thm:SingleUser}}
	\label{app:Proof1}	
For a fixed weighting matrix ${\bf Q}$, the  digital precoding vector ${\bf w}$ that maximizes  \eqref{eq:optimization_problem_single} is 
 \begin{equation} \label{eq: MRT}
     {\bf w}^*=\sqrt{P_{\rm max}}\frac{\left({\bf a}^H\, \mathbf{H} \mathbf{Q}\right)^H}{\left\|{\bf a}^H\, \mathbf{H} \mathbf{Q}\right\|}.
 \end{equation}
 This implies that the maximal ratio transmission with maximum available power is the optimal digital  precoding vector for any fixed ${\bf Q}$.

 By substituting \eqref{eq: MRT} into \eqref{eq:optimization_problem_single} with  $\mathcal{Q}$ replaced by $ \mySet{F}$,  we obtain the optimization problem
 \begin{equation} \label{eq:single_equivalent} 
 \max_{{\bf Q}}~~\left\|{\bf a}^H\, \mathbf{H} \mathbf{Q}\right\|^{2}, \quad  s.t.~~~~\eqref{eq: weighting_matrix},\quad q_{i, l} \in \mySet{F},~ \forall i,l. 
 \end{equation} 
 To drop the non-convex structure constraint on ${\bf Q}$, we rewrite the objective function of \eqref{eq:single_equivalent} as $\left\|{\bf a}^H\, \mathbf{H} \mathbf{Q}\right\|^{2}=\sum_{i=1}^{N_d} \left|\sum_{l=1}^{N_e}A_{i,l}(\p_m)\, e^{ -\jmath k |\p_m-\p_{i,l}|}h_{i,l}q_{i,l}\right|^2$.  
 By substituting this into \eqref{eq:single_equivalent}, we obtain
 \begin{equation} \label{eq:single_equivalent_1x}
 \max_{\left\{q_{i,l}\right\}}~~\sum_{i=1}^{N_d} \left| \sum_{l=1}^{N_e} A_{i,l}(\p_m)\, e^{ -\jmath k |\p_m-\p_{i,l}|}h_{i,l}q_{i,l}\right|^2, \quad s.t.~~~q_{i, l} \in \mySet{F},~ \forall i,l.
 \end{equation} 
 
The maximization problem in \eqref{eq:single_equivalent_1x}  can be decomposed into $N_d$ subproblems, each with an identical structure.  In particular, each subproblem individually designs the weighting coefficients of a single microstrip,  with the $i$th  subproblem given by
 \begin{equation} \label{eq:subproblem} \max_{\left\{q_{i,l}\right\}}~~ \left| \sum_{l=1}^{N_e} A_{i,l}(\p_m)\, e^{ -\jmath k |\p_m-\p_{i,l}|}h_{i,l}q_{i,l}\right|^2, \quad  s.t.~~~q_{i, l} \in \mySet{F},~ \forall l. 
 \end{equation} 
Substituting the expression of $h_{i,l}$   in \eqref{eq:delay_in_microstrip} into \eqref{eq:subproblem}, we obtain
\begin{equation} \label{eq:subproblem_new_exponential_delay}
\max_{\left\{\psi_{i,l}\right\}}~~ \left| \sum_{l=1}^{N_e} A_{i,l}(\p_m)\, e^{-\alpha_i \rho_{i,l}}\, e^{ -\jmath k |\p_m-\p_{i,l}|} e^{-j \beta_i \rho_{i,l}}e^{ \jmath \psi_{i,l}}\right|^2 \,.
\end{equation} 
Hence, according to the triangle inequality, the solution to problem \eqref{eq:subproblem_new_exponential_delay} is
$\psi_{i,l}^*=k|\p_m-\p_{i,l}| + \beta_i \rho_{i,l}$, for each $i,l$, thus proving the theorem.

	%
	\vspace{-0.2cm}
	\subsection{Proof of Theorem \ref{thm:multi-user}}
	\label{app:Proof2}	
	By using the fact that ${\bf x}^T{\bf Q} {\bf y}=({\bf y}^T \otimes {\bf x}^T) {\rm Vec}(\bf Q)$ is valid for arbitrary vectors $\bf x$, $\bf y$, and matrix $\bf Q$, we have 
\begin{equation} \label{eq:reformulate_objective_Q}
    \left|{\bf a}_m^H\, \mathbf{H} \mathbf{Q}\, {\bf w}_m\right|^{2} = \left|({\bf w}_m^T \otimes ({\bf a}_m^H\, \mathbf{H})) {\rm Vec}(\bf Q)\right|^{2},
\end{equation}
\begin{equation} \label{eq:reformulate_objective_Q_j}
    \left|{\bf a}_m^H\, \mathbf{H} \mathbf{Q}\, {\bf w}_j\right|^{2} = \left|({\bf w}_j^T \otimes ({\bf a}_m^H\, \mathbf{H})) {\rm Vec}(\bf Q)\right|^{2}.
\end{equation}

For brevity, we define $L= N_d^2 \times N_e$. Letting ${\bf z}_{m,m}=\left({\bf w}_m^T \otimes ({\bf a}_m^H\, \mathbf{H})\right)^H \in {\mathbb{C}}^{L \times 1}$, ${\bf z}_{j,m}=\left({\bf w}_j^T \otimes ({\bf a}_m^H\, \mathbf{H})\right)^H \in {\mathbb{C}}^{L \times 1}$, and ${\bf q}={\rm Vec}\left(\bf Q \right)\in {\mathbb{C}}^{L \times 1}$, 
we can reformulate the problem \eqref{eq:optimization_problem_DMA} with fixed $\{{\bf w}_m\}$ as
\begin{equation} \label{eq:subproblem_Q_reformulated}
\begin{split}
&\max_{ {\bf q}}~\sum_{m=1}^{M} {\rm log}_{2}\left(1+\frac{\left|{\bf q}^H{\bf z}_{m,m} \right|^{2}}{\sum_{j \neq m}\left|{\bf q}^H{\bf z}_{j,m}\right|^{2}+\sigma^{2}}\right)\\
&s.t.~\left\{\begin{array}{ll}
\left|q_{i}\right| = 1 & {\rm if }~ c_d^i N_e \!+\! 1 \leq c_m^i \leq  (c_d^i \!+\! 1)N_e\\
0 & {\rm otherwise }
\end{array}\right. \quad \forall i=1,\cdots,L,
\end{split}
\end{equation} 
where $c_d^i={\rm div}(i,N_dN_e)$ and $c_m^i = {\rm mod}(i,N_dN_e)$ are the quotient and remainder of the division of $i$ by $N_dN_e$, respectively.
 
 It is easy to verify that the zero elements of ${\bf q}$ have no effect on the objective function of \eqref{eq:subproblem_Q_reformulated}. Hence, \eqref{eq:subproblem_Q_reformulated} can be simplified as
\begin{equation} \label{eq:simplified} 
\max_{ {\bf \bar q}}~~f\left({\bf \bar q}\right) = \sum_{m=1}^{M} {\rm log}_{2}\left(1+\frac{\left|{\bf \bar q}^H{\bf \bar z}_{m,m} \right|^{2}}{\sum_{j \neq m}\left|{\bf \bar q}^H{\bf \bar z}_{j,m}\right|^{2}+\sigma^{2}}\right), \quad s.t.~~~{\bar q_{l}} \in \mathcal{Q},~\forall l \in  \mathcal{A}_q, 
\end{equation} 
where $\mathcal{A}_q$ denotes the set of all non-zero elements of ${\bf q}$, ${\bf \bar q}$ is the modified version of ${\bf q}$ obtained by removing all the zero elements of  ${\bf q}$; and ${\bf \bar z}_{m,m}$ and  ${\bf \bar z}_{j,m}$ are respectively the modified versions of  ${\bf  z}_{m,m}$ and  ${\bf z}_{j,m}$, which are obtained by removing the elements having the same index as the zero elements of ${\bf q}$.
	
\end{appendix}	
\fi 
 
	\bibliographystyle{IEEEtran}
	\bibliography{IEEEabrv,refs}

\end{document}